\documentclass[superscriptaddress,aps,preprintnumbers,amsmath,amssymb,nofootinbib]{revtex4}

\pdfoutput=1

\usepackage{epsfig}
\usepackage{amsmath}
\usepackage{amssymb}
\usepackage{subfigure}
\usepackage{cancel}
\usepackage{textcomp}
\usepackage{calc}
\usepackage{graphicx}
\usepackage{dcolumn}
\usepackage{color}
\usepackage{xcolor}
\usepackage[utf8]{inputenc}

\usepackage{hyperref}
\hypersetup{colorlinks=true,urlcolor=blue,linkcolor=red,citecolor=green!60!black}

\allowdisplaybreaks

\catcode`\@=11
\def\lsim{\mathrel{\mathpalette\@versim<}}
\def\gsim{\mathrel{\mathpalette\@versim>}}
\def\@versim#1#2{\vcenter{\offinterlineskip
\ialign{$\m@th#1\hfil##\hfil$\crcr#2\crcr\sim\crcr } }}
\catcode`\@=12

\newcommand{\al}[1]{\begin{align}#1\end{align}}
\newcommand{\bp}{\begin{pmatrix}}
\newcommand{\ep}{\end{pmatrix}}

\newcommand{\bs}[1]{\boldsymbol}

\newcommand{\be}{\begin{eqnarray}}
\newcommand{\ee}{\end{eqnarray}}

\begin{document}


\title{Planck mass and inflation as consequences of dynamically broken scale invariance}


\author{Jisuke \surname{Kubo}}
\email{kubo@mpi-hd.mpg.de}
\affiliation{Max-Planck-Institut f\"ur Kernphysik (MPIK), Saupfercheckweg 1, 69117 Heidelberg, Germany}
\affiliation{Department of Physics, University of Toyama, 3190 Gofuku, Toyama 930-8555, Japan}

\author{Manfred \surname{Lindner}}
\email{lindner@mpi-hd.mpg.de}
\affiliation{Max-Planck-Institut f\"ur Kernphysik (MPIK), Saupfercheckweg 1, 69117 Heidelberg, Germany}

\author{Kai \surname{Schmitz}}
\email{kai.schmitz@pd.infn.it}
\affiliation{Dipartimento di Fisica e Astronomia, Universit\`a degli Studi di Padova, Via Marzolo 8, 35131 Padova, Italy}
\affiliation{Istituto Nazionale di Fisica Nucleare (INFN), Sezione di Padova, Via Marzolo 8, 35131 Padova, Italy}

\author{Masatoshi \surname{Yamada}}
\email{m.yamada@thphys.uni-heidelberg.de}
\affiliation{Institut f\"ur Theoretische Physik, Universit\"at Heidelberg, Philosophenweg 16, 69120 Heidelberg, Germany}


\begin{abstract}
Classical scale invariance represents a promising framework for model
building beyond the Standard Model.
However, once coupled to gravity, any scale-invariant microscopic model
requires an explanation for the origin of the Planck mass.
In this paper, we provide a minimal example for such a mechanism and show how
the Planck mass can be dynamically generated in a strongly coupled gauge sector.
We consider the case of hidden $SU\left(N_c\right)$ gauge
interactions that link the Planck mass to the condensation
of a scalar bilinear operator that is nonminimally coupled to curvature.
The effective theory at energies below the Planck mass contains
two scalar fields: the pseudo-Nambu--Goldstone boson of spontaneously
broken scale invariance (the dilaton) and a gravitational scalar
degree of freedom that originates from
the $R^2$ term in the effective action (the scalaron).
We compute the effective potential for the coupled dilaton-scalaron system
at one-loop order and demonstrate that it can be used to
successfully realize a stage of slow-roll inflation in the early Universe.
Remarkably enough, our predictions for the primordial scalar and tensor power
spectra interpolate between those of standard $R^2$ inflation
and linear chaotic inflation. 
For comparatively small gravitational couplings, we thus
obtain a spectral index $n_s \simeq 0.97$ and a tensor-to-scalar ratio as large
as $r \simeq 0.08$.
\end{abstract}


\date{\today}
\maketitle
{\hypersetup{linkcolor=black}\tableofcontents}


\section{Introduction}


What is the origin of the Planck mass?
This fundamental question, which we will address in this paper,
attracted much attention in the past, particularly in field
theory~\cite{Fujii:1974bq,Minkowski:1977aj,Zee:1978wi,Terazawa:1981ga,Englert:1975wj,
Englert:1976ep,Chudnovsky:1976zj,Fradkin:1978yw,Smolin:1979uz,Zee:1979hy,Nieh:1982nb,
Terazawa:1976eq,Akama:1977nw,Akama:1977hr,Adler:1980bx,Adler:1980pg,Zee:1980sj,Adler:1982ri}.
If we start with a theory that contains dimensionful parameters,
we cannot explain its origin.
Thus, within the framework of Einstein's theory of gravity, the origin
of the Planck mass cannot be explained.
In the references cited above, conformal symmetry is imposed,
as the Planck mass and hence Einstein gravity can arise through its breaking.
Conformal symmetry can be global~\cite{Fujii:1974bq,Minkowski:1977aj,Zee:1978wi,Terazawa:1981ga}
or local~\cite{Englert:1975wj,Englert:1976ep,Chudnovsky:1976zj,Fradkin:1978yw,
Smolin:1979uz,Zee:1979hy,Nieh:1982nb,Terazawa:1976eq,Akama:1977nw,Akama:1977hr,Adler:1980bx,
Adler:1980pg,Zee:1980sj,Adler:1982ri}, and the theory can contain fundamental
scalar fields from the beginning~\cite{Fujii:1974bq,Minkowski:1977aj,Zee:1978wi,Terazawa:1981ga,
Englert:1975wj,Englert:1976ep,Chudnovsky:1976zj,Fradkin:1978yw,
Smolin:1979uz,Zee:1979hy,Nieh:1982nb}, as in the Brans--Dicke theory~\cite{Brans:1961sx},
or no scalar field as in the case of induced
gravity~\cite{Terazawa:1976eq,Akama:1977nw,Akama:1977hr,
Adler:1980bx,Adler:1980pg, Zee:1980sj,Adler:1982ri}.
Conformal gravity, i.e., gravity supplemented by a local conformal symmetry,
has also been strongly motivating because of its possible renormalizability
(see, e.g.,~\cite{Adler:1982ri,tHooft:2011aa,Mannheim:2011ds}).
If conformal symmetry is imposed in the presence of a scalar field and if it
is an anomaly-free local symmetry, the  scalar field can be eliminated by a
gauge fixing~\cite{Englert:1975wj,Englert:1976ep,Chudnovsky:1976zj,
Fradkin:1978yw,Smolin:1979uz,Zee:1979hy,Nieh:1982nb}.
Alternatively, a certain boundary
condition may be responsible for the generation of Einstein
gravity~\cite{Fradkin:1978yw,Maldacena:2011mk}.
For recent work on the possible connection between conformal gravity
and the origin of the electroweak scale, see also~\cite{Oda:2018zth}.
In contrast to this, if conformal symmetry is global and
spontaneously broken, the Nambu--Goldstone (NG) boson associated
with the breaking appears as a physical degree of freedom (DOF).
This bosonic DOF may be present from the start and
become the NG boson after symmetry breaking.

 
Bosonic DOFs are welcome because they may play the
role of the inflaton field, which is a crucial
element~\cite{Linde:1981mu,Linde:1982zj,Albrecht:1982wi} to overcome
the problems of old inflation~\cite{Guth:1980zm}
(see, e.g.,~\cite{Linde:2007fr} for a historical account).
Since the inflaton field should roll down very slowly at first, the scalar
potential for the inflaton field needs to satisfy the so-called
slow-roll conditions (see, e.g.,~\cite{Baumann:2009ds,Martin:2013tda} for reviews).
Moreover, the shape of the potential is these days very restricted
in order to agree with the PLANCK observations~\cite{Akrami:2018odb}.
Symmetry principles such as the concept of conformal symmetry can help to explain
the origin of the inflaton field in combination with its specific potential.
It can also help if the inflaton field is absent at the beginning as in the case
of $R^2$  inflation~\cite{Starobinsky:1980te,Mukhanov:1981xt,Starobinsky:1983zz} 
(see also \cite{Kehagias:2013mya} and references therein) or if the scalar field
plays a dual role as in the case of Higgs inflation~\cite{Bezrukov:2007ep}
(see also \cite{Rubio:2018ogq} and references therein).


Motivated by these observations and guided by the hints above,
we consider in this paper a model with global conformal symmetry,
which is spontaneously broken by the strong dynamics
in a hidden $SU(N_c)$ gauge theory.
The model contains a complex scalar field $S$ in the fundamental
representation of $SU(N_c)$ with the curvature portal
coupling $S^\dag S\,R$, where $R$ represents the Ricci curvature scalar.
Due to the strong dynamics, a nonzero scalar bilinear
condensate forms, $\langle S^\dag S\rangle\neq 0$, which breaks
conformal symmetry spontaneously~\cite{Osterwalder:1977pc,Fradkin:1978dv}.
The Planck mass is generated dynamically in this way, and the
lowest excitation around the symmetry-breaking
condensate can be identified with the NG boson, i.e., the dilaton.
Note that this mechanism eliminates the generic asymmetry between gravity
with a built-in scale and quantum field theory, where overall scales
have no meaning.
The breaking of conformal symmetry in our scenario sets the scale
in both sectors in a symmetric way.
The potential for the dilaton field and its coupling to the gravitational
field is generated at the same time in this scenario.
Through an appropriate Weyl transformation, the resulting action
can be brought into the Einstein frame.
In this way, we are able to identify the origin of the Planck mass and the inflaton;
it is a mixture of the dilaton and the scalaron, which appears due to the $R^2$ term.

 
Our basic idea to generate the Planck mass and at the same time 
to induce inflation is similar to that of~\cite{Kannike:2015apa}
in that our construction is based on global scale invariance in
the starting classical theory.%
\footnote{References~\cite{GarciaBellido:2011de,Farzinnia:2015fka,Rinaldi:2015uvu,
Ferreira:2016wem,Ghilencea:2018thl,Benisty:2018fja,Barnaveli:2018dxo}
are a partial list of articles that discuss inflation in
classically scale-invariant models (see also the references in these articles).
None of them uses strong dynamics in a non-Abelian gauge theory
to break scale invariance.
Nonperturbative chiral symmetry breaking to produce a robust energy
scale in a classically scale-invariant hidden
sector~\cite{Hur:2011sv,Heikinheimo:2013fta,Holthausen:2013ota}
has been applied to an inflation model~\cite{Inagaki:2015eza}, 
but scale invariance is explicitly broken in their starting classical theory.}
However, our concept to arrive at  the Einstein-Hilbert kinetic term
for the gravitational field that couples to the inflaton field is different:
We rely on the strong dynamics in a non-Abelian gauge theory
that break scale invariance spontaneously, while the model of~\cite{Kannike:2015apa}
has no strongly interacting gauge sector, so the hard breaking of
conformal symmetry by the conformal anomaly~\cite{Callan:1970yg,Symanzik:1970rt,Capper:1974ic}
(i.e., the running of coupling constants) plays a crucial role.
Finally, we mention that the supersymmetric models discussed
in~\cite{Harigaya:2012pg,Harigaya:2014sua,Harigaya:2014wta,Schmitz:2016kyr,
Domcke:2017xvu,Harigaya:2017jny,Domcke:2017rzu} also make use of
strongly coupled gauge dynamics in a hidden sector to generate
the energy scale of inflation.
However, these models simply assume the presence of the Einstein-Hilbert
term from the very beginning and hence offer no dynamical explanation for
the origin of the Planck scale.


Our starting  tree-level action in Eq.~\eqref{LCG2} below has the the most general
form consistent with general diffeomorphism invariance,
$SU(N_c)$ local gauge invariance and global scale invariance.
Then, along the lines of the Nambu--Jona-Lasinio (NJL)
model~\cite{Nambu:1961tp,Nambu:1961fr}, we proceed to an effective-theory
description of the scalar bilinear condensate.
In the mean-field approximation, we are not only able to identify the
dilaton but can also derive its potential by integrating out the
fluctuations of the scalar field $S$ around its background.
In doing so, we arrive at the one-loop effective action, which is the
starting action in the Jordan frame in order to subsequently discuss
inflation in section III.
In this part of our analysis, we assume that the Weyl tensor squared is
negligible for the purposes of inflation, which is why we
suppress it in the effective action.
The final effective Lagrangian in the Einstein frame involves the scalaron
field as well.
We find that the scalar potential is such that the coupled dilaton-scalaron
system gives rise to an effective single-field model of inflation.
In the last part of section III, we perform a numerical analysis
to compute the inflationary observables encoded in the cosmic microwave
background (CMB) and to compare our predictions with
the latest data from the PLANCK satellite mission~\cite{Akrami:2018odb}.
Our model contains four independent parameters, two of which are used to fix
the values of the reduced Planck mass $M_{\rm Pl}$ and the amplitude $A_s$ of
the primordial scalar power spectrum.
We are hence left with two free
parameters to  describe the inflationary Universe.
The last section is devoted to conclusions.
 

\section{Dynamical origin of the Planck mass}
\label{sec:model}


\subsection{Spontaneous breaking of scale invariance in a hidden strongly coupled sector}


The starting point of our analysis is the most general action that complies
with the symmetry principles of general diffeomorphism invariance,
local gauge invariance, and global scale invariance at the classical level,
\al{
S_\text{C} = \int d^4 x\,\sqrt{-g} \left[
- \hat{\beta}\,S^\dag S\,R +\hat{\gamma}\,R^2 - \frac{1}{2}\,\mbox{Tr}\,F^2
+ g^{\mu\nu}\left(D_\mu S\right)^\dag D_\nu S - \hat{\lambda}\,(S^\dag S)^2
+ a\,R_{\mu\nu}R^{\mu\nu} + b\,R_{\mu\nu\alpha\beta}R^{\mu\nu\alpha\beta}
\right] \,.
\label{LCG2}}
Here, $R$ denotes the Ricci curvature scalar, $R_{\mu\nu}$ is the Ricci tensor,
$R_{\mu\nu\alpha\beta}$ is the Riemann tensor, $F$ is the matrix-valued
field-strength tensor of the $SU(N_c)$ gauge theory, and $S$ is the
complex scalar field in the fundamental representation of
$SU(N_c)$, with $D_\mu$ being the covariant derivative.
Note that the action $S_C$ also features an accidental global $U(1)$
symmetry corresponding to $S$ number conservation.
The last two terms in Eq.~\eqref{LCG2} can be written as
\al{
\left(\frac{1}{2}\,a + 2\,b\right) W_{\mu\nu\alpha\beta}W^{\mu\nu\alpha\beta} -
\left(\frac{1}{2}\,a + b\right) E + \frac{1}{3}(a+b)R^2 \,,
\label{Weyl1}}
where $W_{\mu\nu\alpha\beta} = R_{\mu\nu\alpha\beta}+
\frac{1}{2}\left(-g_{\mu\alpha}R_{\nu\beta}+g_{\mu\beta}R_{\nu\alpha} +
g_{\nu\alpha}R_{\mu\beta} - g_{\nu\beta}R_{\mu\alpha}\right)+ \frac{1}{6}
\left(g_{\mu\alpha}g_{\nu\beta}-g_{\nu\alpha}g_{\mu\beta}\right)R$ denotes the Weyl tensor
and $E = R_{\mu\nu\alpha\beta}R^{\mu\nu\alpha\beta} - 4\,R_{\mu\nu}R^{\mu\nu} + R^2$
is the Gau\ss-Bonnet term, which we will suppress because the corresponding
action is a surface term.
The $R^2$ term in Eq.~\eqref{Weyl1} can be absorbed into a
redefinition of the coefficient $\gamma$, so the last two terms
in Eq.~\eqref{LCG2} can be replaced by a term of the form
$\kappa\,W_{\mu\nu\alpha\beta}W^{\mu\nu\alpha\beta}$, which,
multiplied with $\sqrt{-g}$, is invariant under local conformal
transformations, $g_{\mu\nu}(x)\to g'_{\mu\nu}(x)=\Omega^2(x)\,g_{\mu\nu}(x)$.


In our scenario~\cite{Kubo:2014ova,Kubo:2015cna,Kubo:2015joa}, the
condensation of the gauge-invariant scalar bilinear $\langle S^\dag S\rangle$
in the confining phase breaks scale symmetry spontaneously, in a
way similar to how the chiral fermion condensate breaks chiral symmetry in QCD.%
\footnote{A gauge boson condensate $\langle \textrm{Tr}\,F^2\rangle$
can also form in the confining phase.
This condensate breaks scale invariance dynamically and can be used to generate
the Planck mass~\cite{Zee:1980sj,Adler:1982ri,Salvio:2017qkx}.
In this paper, we shall ignore this dynamical breaking of scale invariance.}
However, a consistent formulation of the quantum theory for the classical
action in Eq.~\eqref{LCG2} is not yet available.
One possibility of such a formulation utilizes the notion of asymptotic
safety~\cite{weinberg} (see, e.g., \cite{Eichhorn:2018yfc} and references therein),
which assumes the existence of a nontrivial ultraviolet fixed point.
At the ultraviolet fixed point, conformal symmetry is restored as an unbroken
symmetry of the full quantum theory.
In the following discussion, we are implicitly assuming that the hard breaking
of scale invariance by the scale anomaly is weak enough, so we may ignore it
for the purposes of describing the spontaneous breaking of scale invariance.
Related to that, we recall Ref.~\cite{Bardeen:1985sm}, 
in which it has been pointed out that spontaneous scale symmetry breaking
can be  associated with the chiral condensate in QCD if the running of the
gauge coupling is slow (i.e., if the hard breaking by the scale anomaly is weak).
To realize a slow running of the gauge coupling 
as well as of the quartic coupling $\hat{\lambda}$ in our model,
a certain set of fermion fields should be introduced~\cite{Cheng:1973nv}.
But we will not go into such details, which are beyond the scope of the present work.


Even if gravity is switched off, it is  nontrivial to investigate whether or
not a scalar condensate really forms~\cite{Osterwalder:1977pc,Fradkin:1978dv}.
In the following, instead of investigating this problem from first principles, we
will rely on an effective theory for the order parameter $\langle S^\dag S\rangle$
(strictly speaking, it is an approximate order
parameter~\cite{Osterwalder:1977pc,Fradkin:1978dv}).
Let us outline the basics of the effective theory that 
has been proposed in~\cite{Kubo:2015cna,Kubo:2015joa} by applying the
concepts of the NJL model~\cite{Nambu:1961tp,Nambu:1961fr}, which is an
effective theory for the chiral condensate, i.e., the order parameter
of chiral symmetry breaking in QCD.
As in the case of the NJL model, the effective Lagrangian does not
contain the gauge fields because they are integrated out, while it
contains the ``constituent'' scalar field $S$, similar to the
constituent quarks in the NJL model,
\al{
S_\text{C,eff} = \int d^4 x\,\sqrt{-g} \left[
- \beta\,S^\dag S\,R+ \gamma\,R^2
+ \kappa\,W_{\mu\nu\alpha\beta} W^{\mu\nu\alpha\beta}
+ g^{\mu\nu}\left(\partial_\mu S\right)^\dag \partial_\nu S
- \lambda\,\left(S^\dag S\right)^2\right] \,.
\label{LCGeff2}}
This is again the most general action that is consistent with general diffeomorphism
invariance, global $SU(N_c)\times U(1)$ symmetry, and classical global scale invariance.%
\footnote{An attempt at taking into account the effect of confinement
was made in~\cite{Kubo:2018vdw}.}
In writing down Eq.~\eqref{LCGeff2}, we omitted all higher-dimensional operators
arising from integrating out the gauge fields that explicitly break scale invariance.
These operators reflect the explicit breaking of scale invariance at the quantum
level, which we assume to be subdominant compared to the spontaneous breaking of
scale invariance by the scalar condensate at low energies.
The couplings $\hat{\beta}$, $\hat{\gamma}$, and $\hat{\lambda}$
in the fundamental action $S_\text{C}$ are not the same as $\beta$, $\gamma$,
and $\lambda$ in the effective action $S_{\rm C, eff}$ because the latter are
effective couplings that are dressed by gauge field contributions.
Introducing the auxiliary field (mean field) $f$, we can rewrite
Eq.~\eqref{LCGeff2} as
\al{
S_\text{C,eff} = \int d^4 x\,\sqrt{-g} \left[
\gamma\,R^2 + \kappa\,W_{\mu\nu\alpha\beta }W^{\mu\nu\alpha\beta}
+ g^{\mu\nu}\left(\partial_\mu S\right)^\dag \partial_\nu S
-\left(2\,\lambda\,f + \beta\,R\right)S^\dag S + \lambda\,f^2 \right] \,.
\label{LCGeff3}}
The action in Eq.~\eqref{LCGeff2} can be obtained from Eq.~\eqref{LCGeff3} by
inserting the equation of motion for the auxiliary field $f$,
\al{
f=S^\dag S\,.
\label{eq_f}}
Next, let us compute the one-loop effective potential for the field $f$.
Scalar and fermionic loop corrections to the vacuum energy density in a curved (de Sitter)
background were computed for the first time in~\cite{Candelas:1975du}
(see also \cite{DeWitt:1975ys}).
To obtain the effective potential for $f$, we integrate out 
the fluctuations of the field $S$ around its background value $\bar{S}$,
\al{
V_{\rm eff}\left(f,\bar{S},R\right) =
\left(2\,\lambda\,f + \beta\,R\right) \bar{S}^\dag \bar{S}
- \lambda\,f^2
+ \frac{N_c}{32 \pi^2}\left(2\,\lambda\,f + \beta\,R\right)^2
\,\ln\frac{2\,\lambda\,f + \beta\,R}{\Lambda^2} \,,
\label{Vb}}
where $\Lambda=e^{3/4}\,\mu_{\overline{\rm MS}}$.
We used dimensional regularization together with renormalization
according to the modified minimal subtraction scheme,%
\footnote{The Gaussian path integral has been performed assuming
a flat spacetime metric.
If the fluctuations around the flat metric are taken into account, a term
$\propto \sqrt{-g}\left(R_{\mu\nu}R^{\mu\nu}-R_{\mu\nu\alpha\beta}R^{\mu\nu\alpha\beta}\right)$
as well as an additional term $\propto  \sqrt{-g}\,R^2$ will be generated
at one-loop order~\cite{tHooft:1974toh,Casarin:2018odz}.
They can be absorbed into redefinitions of $\kappa$ and $\gamma$
in Eq.~\eqref{LCGeff3}, respectively, if the topological $E$ term is ignored.}
where the divergences are absorbed in the coupling constants as follows,
$\lambda \to \lambda - 4\,\lambda^2 L$,
$\beta \to \beta - 4\,\beta\,\lambda\,L$,
$\gamma \to \gamma + \beta^2 L$,
together with the shift
$f \to f + \left(4\,\lambda\,f + 2\,\beta\,R\right)L$, where
$L = \left(1/\epsilon + \ln 4\pi - \gamma_E\right)N_c/32 \pi^2$.


To compute the expectation value $\langle f\rangle $ of the mean field $f$, we assume a negligibly small, but nonzero value of the curvature scalar, $\langle R\rangle \ll \langle f \rangle$.
During the (quasi) de Sitter stage of expansion during inflation, the curvature scalar will be determined by the inflationary Hubble rate, $\langle R \rangle \simeq 12\,H_{\rm inf}^2$, up to slow-roll corrections.
This relation will allow us to justify the assumption $\langle R\rangle \ll \langle f \rangle$ \textit{a posteriori} based on the parametric separation of the energy scales $\langle f \rangle^{1/2}$ and $H_{\rm inf}$.
In practice, we are now going to compute the expectation value $\langle f\rangle $ in two steps.
First, we expand the effective potential $V_{\rm eff}$ in Eq.~\eqref{Vb} in powers of the small dimensionless ratio $\left(\beta R\right)/\left(\lambda f\right)$; then we minimize the $R$-independent leading-order contribution to this expansion with respect to the fields $f$ and $\bar{S}$.
This results in
\al{
\langle f\rangle = f_0 = \frac{\Lambda^2}{2\,\lambda}\,
\exp\left(\frac{8\pi^2}{N_c\lambda}-\frac{1}{2}\right)
\quad\textrm{and}\quad \langle\bar{S}\rangle = 0 \,.
\label{f0}}
Our result for $f_0$ is controlled
by the one-loop contribution to the effective potential.
This may raise the concern that $f_0$ in Eq.~\eqref{f0} actually lies outside the
range of validity of the effective potential.
At this point, it is, however, important to recognize that $V_{\rm eff}$ in Eq.~\eqref{Vb}
is the outcome of a nonperturbative calculation based on the mean-field approximation.
Within this approximation, Eq.~\eqref{f0} therefore represents a consistent
expression for the expectation value of the mean field $f$.
In fact, the expression for $f_0$ in Eq.~\eqref{f0} is a central result of our analysis:
\textit{In view of the action in Eq.~\eqref{LCGeff2}, together with
Eq.~\eqref{eq_f}, we now see that the strong dynamics in the non-Abelian
gauge theory can break conformal symmetry spontaneously and consequently
generate the Planck mass dynamically,}
\al{
M_\text{Pl}^2 = 2\,\beta\,f_0 = \frac{N_c\beta}{16\pi^2} \left(2\,\lambda\,f_0\right)
\left(1 + 2\,\ln\frac{2\,\lambda\,f_0}{\Lambda^2}\right) \,,
\label{planck}}
where  $M_{\rm Pl}$ is the reduced Planck mass.
At the same time, the constituent scalar field does not acquire a nonvanishing
vacuum expectation value, $\left<\bar{S}\right> = 0$.
The global $U(1)$ symmetry associated with the conservation of $S$ particle number
therefore remains unbroken.


At the minimum in Eq.~\eqref{f0} (and still neglecting higher-order terms in $R$), the expectation value of $V_{\rm eff}$ is given by
\al{
\langle V_{\rm eff}\rangle = -\frac{N_c}{64\pi^2}\left(2\,\lambda\,f_0\right)^2 =
- \frac{N_c\lambda^2}{64\pi^2\,\beta^2}\,M^4_{\rm Pl} = - U_0\,.
\label{CC}}
The spontaneous breaking of scale symmetry thus generates 
a negative vacuum energy, which is $U_0 = -8\times 10^{-9}M^4_{\rm Pl}$
for $\lambda=1$, $N_c=5$, $\beta=10^3$.
The zero-point energy density $U_0$ is finite in dimensional regularization
because of the scale invariance of the action in Eq.~\eqref{LCGeff3}.
If we use another regularization scheme, it can be different or even divergent.
This reflects the fact that the zero-point energy density cannot be uniquely
determined within the framework of quantum field theory in flat spacetime:
The  cosmological constant problem remains still unsolved,
although we established a link between particle physics and gravity
via spontaneous breaking of global conformal symmetry.
Here we are not attempting to solve this problem and proceed 
with our discussion in the hope that there will be a mechanism
in, e.g., quantum gravity that solves this problem.
In the following discussion, we thus simply subtract this vacuum energy
density from the potential and continue with the potential
$V_{\rm eff } \to V_{\rm eff } + U_0$
so that the potential has zero cosmological constant.
Consequently, the effective potential at the minimum (now restoring the $R$ dependence) can be written as
\al{
\label{Veffhat}
V_{\rm eff}\left(f_0,0,R\right) & = U_0 - \lambda\,f_0^2
+ \frac{N_c}{32\pi^2}\left(2\,\lambda\,f_0\right)^2\,\ln\frac{2\,\lambda\,f_0}{\Lambda}
\\ \nonumber
& + \frac{N_c}{32\pi^2}(2\,\lambda\,f_0)
\left(1 + 2\,\ln\frac{2\,\lambda\,f_0}{\Lambda^2}\right)\beta\,R 
+ \frac{N_c}{64\pi^2}\left(3 + 2\,\ln\frac{2\,\lambda\,f_0}{\Lambda^2}\right)\beta^2 R^2
+ \mathcal{O}\left(\frac{\beta^3 R^3}{2\,\lambda\,f_0}\right)\,,}
from which we see that the Planck mass is correctly identified
in Eq.~\eqref{planck} and that the parameter $\gamma$ is shifted to
\al{
\bar{\gamma} = \gamma - \frac{\beta^2}{2\lambda}\left(\frac{1}{2}+\ell\right)
\quad\textrm{with}\quad \ell=\frac{N_c \lambda}{16\pi^2} \,,
\label{gamma}}
where Eq.~(\ref{f0}) has been used.
 

\subsection{Effective action for the dilaton field at low energies}


The mean field $f$ is a scalar field with canonical mass dimension two,
and it can describe the excitations above the vacuum expectation value $f_0$.
At the one-loop level, the mean field $f$ obtains a kinetic term,
and the excitations can be described by a propagating real scalar field.
It is important to realize that this result is a characteristic outcome
of our nonperturbative analysis based on the mean-field approximation.
Indeed, studying our effective model in Eq.~\eqref{LCGeff3}
for the original gauge theory in Eq.~\eqref{LCG2} in the mean-field approximation
allows us to identify the mean field $f$ as the propagating dilaton
field, $f \sim \chi^2$ [see Eq.~\eqref{chi} below], which we know
must exist in the confining phase at low energies because of the
spontaneous breaking of scale invariance by the scalar condensate.
We emphasize that this would not be possible if we simply treated our
effective model in Eq.~\eqref{LCGeff3}
according to the rules of conventional perturbation theory.
In this case, the mean field $f$ would remain an auxiliary field, and we would have
no chance of describing the $S^\dagger S$ bound state.
In particular, we would be unable to correctly identify the propagating dilaton
field at low energies.


Suppressing the complex scalar $S$, we now proceed with the effective action
\al{
\Gamma_{\rm eff} = \int d^4 x\,\sqrt{-g}\left[\frac{1}{2}\,Z^{-1}\,
g^{\mu\nu}\partial_\mu f\,\partial_\nu f - V_{\rm eff}\left(f,R\right) +
\gamma\,R^2 + \kappa\,W_{\mu\nu\alpha\beta}W^{\mu\nu\alpha\beta} \right] \,,
\label{Gamma1}}
where
\al{
\label{Veff}
V_{\rm eff}\left(f,R\right) & = U_0 - \lambda f^2 +
\frac{N_c}{32\pi^2}\left(2\,\lambda\,f\right)^2\,\ln\frac{2\,\lambda\,f}{\Lambda^2}
\\ \nonumber
& + \frac{N_c}{32\pi^2}\left(2\,\lambda\,f\right)
\left(1 + 2\,\ln\frac{2\,\lambda\,f}{\Lambda^2}\right)\beta\,R
+ \frac{N_c}{64\pi^2}\left(3 + 2\,\ln\frac{2\,\lambda\,f}{\Lambda^2}\right)\beta^2 R^2
+ \frac{N_c \beta^3 R^3}{96\pi^2\left(2\,\lambda \,f\right)}
+ \mathcal{O}\left(\frac{\beta^4 R^4}{(2\,\lambda\,f)^2}\right)\,.}
The wave function renormalization constant $Z^{-1}$ is given by
\al{
Z^{-1} =
\frac{N_c\lambda^2}{24\pi^2\left(2\,\lambda\,f + \beta\,R\right)}
= \frac{N_c\lambda^2}{24\pi^2\left(2\,\lambda\,f\right)}
\left[1 - \frac{\beta\,R}{2\,\lambda\,f}
+ \frac{\beta^2R^2}{\left(2\,\lambda\,f\right)^2} + \dots\right] \,.}
An explicit derivation of this result for the wave function renormalization
constant can be found in~\cite{Kubo:2016kpb}.
Next, we proceed from $f$ with canonical dimension two to
the canonically normalized field $\chi$ with canonical dimension one, i.e.,
\al{
f = \left(\frac{48\pi^2}{N_c\lambda}\right)(\chi + \chi_0)^2
\quad\textrm{with}\quad
\chi_0 = \left(\frac{N_c\lambda f_0}{48\pi^2}\right)^{1/2}
= \left(\frac{\ell}{6\beta}\right)^{1/2} M_{\rm Pl} \,,
\label{chi}}
where $\ell$ is defined in Eq.~(\ref{gamma}).
We find that the effective action in Eq.~\eqref{Gamma1} can be
written as $\Gamma_{\rm eff} =\int d^4 x\,\mathcal{L}_{\rm eff}$, where
\al{
\frac{\mathcal{L}_{\rm eff}}{\sqrt{-g}} =
- \frac{1}{2}\,M^2_{\rm Pl}\,B\left(\chi\right)\,R
+ G\left(\chi\right) R^2 + \kappa\,W_{\mu\nu\alpha\beta} W^{\mu\nu\alpha\beta}
+ \frac{1}{2}\,g^{\mu\nu}\partial_\mu \chi\,\partial_\nu \chi - U\left(\chi\right)
+ \dots \,.
\label{Gamma2}}
Here, the ellipsis stands for higher-order terms such as
$g^{\mu\nu} \partial_\mu\chi\,\partial_\nu\chi\,\beta\,R/\chi_0^2$,
$\beta^3 R^3/\chi_0^2$, etc.
The dilaton-dependent functions $B\left(\chi\right)$, $G\left(\chi\right)$,
and $U\left(\chi\right)$ are given by
\begin{align}
\label{eq:B}
B\left(\chi\right) & = \left(1 + \frac{\chi}{\chi_0}\right)^2 \left[
1 + 2\,\ell\ln\left(1 + \frac{\chi}{\chi_0}\right)^2\right] \,, \\
\label{eq:Gchi}
G\left(\chi\right) & = \bar{\gamma} - \frac{N_c\beta^2}{32\pi^2}
\,\ln\left(1 + \frac{\chi}{\chi_0}\right)^2 \,, \\
\label{eq:Uchi}
U\left(\chi\right) & = 
U_0 \left(1 + \frac{\chi}{\chi_0}\right)^4 \left[
2\ln\left(1 + \frac{\chi}{\chi_0}\right)^2 - 1\right] + U_0 \,.
\end{align}
$U_0$ and $\bar{\gamma}$ are defined in Eqs.~(\ref{CC}) and (\ref{gamma}),
respectively.
The effective action ${\cal L}_{\rm eff}$ in Eq.~\eqref{Gamma2} illustrates
that the dilaton field $\chi$ is, in fact, not a true NG
boson in the sense that it parametrizes an exactly flat direction in the scalar potential.
Instead, $\chi$ is identified as the pseudo-NG associated with the spontaneous breaking
of classical scale invariance.
The nontrivial interactions of the dilaton field reflect the fact that our effective theory
defined in Eq.~\eqref{LCGeff2} violates scale invariance at the quantum level.
To arrive at ${\cal L}_{\rm eff}$ in Eq.~\eqref{Gamma2}, we used 
Eqs.~\eqref{f0} and \eqref{chi} to find
\al{
\ln\left(\frac{2\,\lambda\,f}{\Lambda^2}\right) =
\ln\left(\frac{96 \pi^2 \chi_0^2}{N_c \Lambda^2}\right)
\left(1+\frac{\chi}{\chi_0}\right)^2 =
\frac{8\pi^2}{N_c\lambda} - \frac{1}{2} + \ln\left(1 + \frac{\chi}{\chi_0}\right)^2 \,.}
$V_{\rm eff}$ in Eq.~\eqref{Veff} is obtained from Eq.~\eqref{Vb}
by expanding in powers of $\beta\,R$.
This expansion makes sense only if $2\,\lambda\,f \gg \beta\,R$, which means
$M^2_{\rm Pl}\left(1 + \chi/\chi_0\right)^2 \gg \left(\beta^2/\lambda\right)R$.
Otherwise, it is impossible to generate Einstein gravity.
Furthermore, even if
$M^2_{\rm Pl}\left(1 + \chi/\chi_0\right)^2 \gg \left(\beta^2/\lambda\right)R$
is satisfied, the coefficient of $R$ in Eq.~\eqref{Gamma2}
still vanishes as soon as $B\left(\chi\right) = 0$.


\section{Dilaton-scalaron inflation}


\subsection{Full scalar potential of the dilaton-scalaron system}


In the previous section, we discussed how the condensation of a scalar bilinear
operator nonminimally coupled to the Ricci scalar leads to the dynamical generation
of the Planck mass.
In this section, we shall now turn to the effective theory at energies below
the Planck scale and demonstrate how the scalar DOFs at low energies
can give rise to a stage of inflationary expansion in the early Universe.
The starting point of our analysis is the effective Jordan-frame Lagrangian
in Eq.~\eqref{Gamma2}, truncated after the second order in the Jordan-frame
Ricci scalar $R_J$,
\begin{align}
\label{eq:LagJ}
\frac{\mathcal{L}_{\rm eff}^J}{\sqrt{- g_J}} =
-\frac{1}{2}\,B\left(\chi\right) M_{\rm Pl}^2\, R_J +
G\left(\chi\right) R_J^2 + \frac{1}{2}\,g_J^{\mu\nu}\partial_\mu\chi\,\partial_\nu \chi
- U\left(\chi\right) \,.
\end{align}
Here, $g_J^{\mu\nu}$ and $g_J$ denote the inverse and the determinant of
the Jordan-frame spacetime metric $g_{\mu\nu}^J$, respectively.
By construction, the kinetic term of the dilaton field $\chi$ is
canonically normalized in the Jordan frame.
In this work, we shall assume that the term involving the Weyl tensor squared
is negligible for the purposes of inflation.
The equation of motion for the dilaton field
is, in any case, independent of this term.
Moreover, we recall that the dimensionless function $B(\chi)$ 
given in Eq.~(\ref{eq:B}) parametrizes
the nonminimal coupling of the dilaton field $\chi$ to the Ricci scalar $R_J$,
while the dimensionless function $G(\chi)$ given in Eq.~(\ref{eq:Gchi})
denotes the field-dependent coefficient of the $R_J^2$ term.
The dimensionful function $U(\chi)$  given in Eq.~(\ref{eq:Uchi})
stands for the Jordan-frame scalar potential.
A similar Lagrangian with \textit{a priori} arbitrary
functions $B$, $G$, and $U$ has been studied in~\cite{Kaneda:2015jma}
for purely phenomenological reasons.


The Lagrangian in Eq.~\eqref{eq:LagJ} admits solutions of the equations of motion
that describe an inflationary stage of expansion.
To see this more clearly, we need to transform Eq.~\eqref{eq:LagJ} from the Jordan
frame to the Einstein frame.
In a first step, let us introduce an auxiliary field $\psi$ with mass dimension two
that allows us to remove the $R_J^2$ term,
\begin{align}
\label{eq:LagJS}
\frac{\mathcal{L}_{\rm eff}^J}{\sqrt{- g_J}} = -\left[
\frac{1}{2}\,B\left(\chi\right) M_{\rm Pl}^2
- 2\,G\left(\chi\right)\psi\right] R_J
+ \frac{1}{2}\,g_J^{\mu\nu}\partial_\mu\chi\,\partial_\nu \chi
- U\left(\chi\right) - G\left(\chi\right) \psi^2 \,.
\end{align}
Indeed, varying this Lagrangian w.r.t.\ the auxiliary field,
$\delta_\psi\,\mathcal{L}_{\rm eff}^J = 0$, results in $\psi = R_J$, which
restores the Lagrangian in Eq.~\eqref{eq:LagJ}.
In a second step, we perform a Weyl rescaling of the metric,
$g_{\mu\nu} = \Omega^2\, g_{\mu\nu}^J$, where we choose the conformal
factor $\Omega$ such that the kinetic term for the gravitational field
obtains its standard Einstein-Hilbert form,
\begin{align}
\label{eq:WeylTrafo}
\Omega^2\left(\chi,\psi\right) =
B\left(\chi\right) - \frac{4\,G\left(\chi\right)\psi}{M_{\rm Pl}^2} \,.
\end{align}
At this point, it is helpful to remember that the Ricci scalar $R_J$ behaves as follows
under a Weyl transformation,
\begin{align}
\label{eq:RJWeyl}
R_J = \Omega^2 \left(R +3\,g^{\mu\nu}\partial_\mu\,\partial_\nu \ln\Omega^2
- \frac{3}{2}\,g^{\mu\nu}\,\partial_\mu \ln\Omega^2\,\partial_\nu \ln\Omega^2 \right) \,.
\end{align}
Neglecting the total derivative in this expression, we thus arrive at the
Einstein-frame Lagrangian
\begin{align}
\label{eq:LagEO}
\frac{\mathcal{L}_{\rm eff}^E}{\sqrt{- g}} = -\frac{1}{2}\,M_{\rm Pl}^2
\left(R - \frac{3}{2}\,g^{\mu\nu}\,\partial_\mu \ln\Omega^2\left(\chi,\psi\right)\,
\partial_\nu \ln\Omega^2\left(\chi,\psi\right)\right)
+ \frac{g^{\mu\nu}}{2\,\Omega^2\left(\chi,\psi\right)}\,\partial_\mu\chi\,\partial_\nu \chi
- V\left(\chi,\psi\right) \,,
\end{align}
where $V$ denotes the scalar potential in the Einstein frame,
\begin{align}
V\left(\chi,\psi\right) =
\frac{U\left(\chi\right) + G\left(\chi\right) \psi^2}{\Omega^4\left(\chi,\psi\right)} =
\frac{U\left(\chi\right) + G\left(\chi\right) \psi^2}
{\left(B\left(\chi\right) M_{\rm Pl}^2 -
4\,G\left(\chi\right)\psi\right)^2} \, M_{\rm Pl}^4 \,.
\end{align}
The Lagrangian in Eq.~\eqref{eq:LagEO} indicates that the function
$\ln\Omega^2$, which we extracted from the Jordan-frame Ricci
scalar $R_J$ in Eq.~\eqref{eq:RJWeyl}, can be regarded as a propagating scalar field
that possesses its own kinetic term.
In the following, we will refer to this new scalar field as the scalaron $\phi$;
the canonically normalized scalaron is defined as follows,
\begin{align}
\label{eq:phi}
\phi = \sqrt{\frac{3}{2}}\,M_{\rm Pl} \ln\Omega^2 \,.
\end{align}
Making use of this definition, the conformal factor $\Omega^2$ can
be written as an exponential function of the scalaron field,
\begin{align}
\Omega^2 = e^{\Phi\left(\phi\right)} \,, \quad \
\Phi\left(\phi\right) = \frac{\sqrt{2}\,\phi}{\sqrt{3}\,M_{\rm Pl}}  \,.
\end{align}
Similarly, thanks to Eqs.~\eqref{eq:WeylTrafo} and \eqref{eq:phi},
the auxiliary field $\psi$ can be traded for a function of the fields $\chi$ and $\phi$,
\begin{align}
\psi = \frac{M_{\rm Pl}^2}{4\,G\left(\chi\right)}\left[B\left(\chi\right)
- e^{\Phi\left(\phi\right)}\right] \,.
\end{align}
The Einstein-frame Lagrangian for the coupled dilaton-scalaron system thus obtains
the form
\begin{align}
\label{eq:LEphichi}
\frac{\mathcal{L}_{\rm eff}^E}{\sqrt{- g}} = -\frac{1}{2}\,M_{\rm Pl}^2\,R
+ \frac{1}{2}\,g^{\mu\nu}\,\partial_\mu\phi\,\partial_\nu \phi 
+ \frac{1}{2}\,e^{-\Phi\left(\phi\right)}\,g^{\mu\nu}\,\partial_\mu\chi\,\partial_\nu \chi
- V\left(\chi,\phi\right)  \,,
\end{align}
where the Einstein-frame scalar potential $V$ as a function of $\chi$ and $\phi$ now reads
\begin{align}
\label{eq:V}
V\left(\chi,\phi\right) = 
e^{-2\,\Phi\left(\phi\right)} \left[ U\left(\chi\right)
+ \frac{M_{\rm Pl}^4}{16\,G\left(\chi\right)}\left(B\left(\chi\right)
- e^{\Phi\left(\phi\right)}\right)^2\right] \,.
\end{align}
This scalar potential can be used as the starting point for the analysis
of slow-roll inflation.


We mention in passing that a scalar potential similar to
the one in Eq.~\eqref{eq:V} has recently also been obtained in~\cite{Karam:2018mft},
which presents a study of Coleman-Weinberg inflation supplemented by a
nonminimal inflaton coupling to gravity and an additional $R^2$ term.
It is therefore intelligible that our analysis shares a number
of similarities with the one in~\cite{Karam:2018mft}.
On the other hand, one should notice that the model in~\cite{Karam:2018mft}
builds upon the assumption of simple polynomial expressions in the initial
Jordan-frame Lagrangian, whereas our functions $B$, $G$, and $U$
in Eq.~\eqref{eq:LagJ} all include logarithmic terms that
are generated radiatively in the low-energy effective action.
This explains why the field dependence of our scalar potential is more complicated,
after all, than in the case of the scalar potential discussed in~\cite{Karam:2018mft}.


\subsection{Effective single-field description along the inflationary trajectory}


The scalar potential in Eq.~\eqref{eq:V} is a complicated function in the two-dimensional
field space spanned by $\chi$ and $\phi$.
In principle, one could therefore imagine that Eq.~\eqref{eq:V} allows one
to realize inflation in various parameter regimes and different parts of field space.
In particular, this might include scenarios in which both scalar fields
act as slowly rolling inflatons, such that inflation needs to be described
by a full-fledged two-field analysis.
In the following, however, we will refrain from attempting to identify
such intricate scenarios.
Instead, we will focus on a particularly simple inflationary solution
that can be described by an effective single-field model.
Our key observation is that the scalar potential in Eq.~\eqref{eq:V} always
possesses exactly one local extremum in the scalaron
direction at $\phi = \phi_\star$, 
\begin{align}
\label{eq:phi0}
\left.\frac{\partial\,V}{\partial\phi}\right|_{\left(\chi,\phi_\star\left(\chi\right)\right)} = 0
\qquad\Rightarrow\qquad
\phi_\star\left(\chi\right) = \sqrt{\frac{3}{2}}\,M_{\rm Pl}
\ln\left[\left(1+ 4\,A\left(\chi\right)\right)B\left(\chi\right)\right] \,,
\end{align}
where we introduced the dilaton-dependent
function $A$, which will allows us to simplify our notation
in the following,
\begin{align}
\label{eq:Adef}
A\left(\chi\right) = \frac{4\,G\left(\chi\right) U\left(\chi\right)}
{B^2\left(\chi\right)M_{\rm Pl}^4} \,.
\end{align}
The presence of this extremum motivates us to seek inflationary solutions
along the following contour in field space,
\begin{align}
\label{eq:contour}
\mathcal{C}\left(\sigma\right) = 
\left\{\chi\left(\sigma\right),\phi\left(\sigma\right)\right\} = 
\left\{\sigma,\phi_\star\left(\sigma\right)\right\} \,,
\end{align}
where $\sigma$ can be regarded as the physical inflaton
field that parametrizes the one-dimensional contour $\mathcal{C}$
in the two-dimensional field space.
Along this trajectory, the dilaton-scalaron system can be effectively
described by a single-field model in terms of the real scalar $\sigma$.
Interestingly enough, a similar observation has recently
also been made in~\cite{Gundhi:2018wyz}, which contains a study of
Higgs inflation in the presence of an extra $R^2$ term.
The model in~\cite{Gundhi:2018wyz} describes a coupled Higgs-scalaron system
that can also be reduced to an effective single-field model along a local extremum
in the scalar potential.
We point out that the model~\cite{Gundhi:2018wyz} simply
assumes the presence  of the Einstein-Hilbert term from the outset.
The particle spectrum of this model therefore does not contain a dilaton
field that could couple to the scalaron; the role of our dilaton field $\chi$
is thus played by the only other available scalar DOF: the standard
model Higgs boson.

\begin{figure}
\begin{center}
\includegraphics[width=0.475\textwidth]{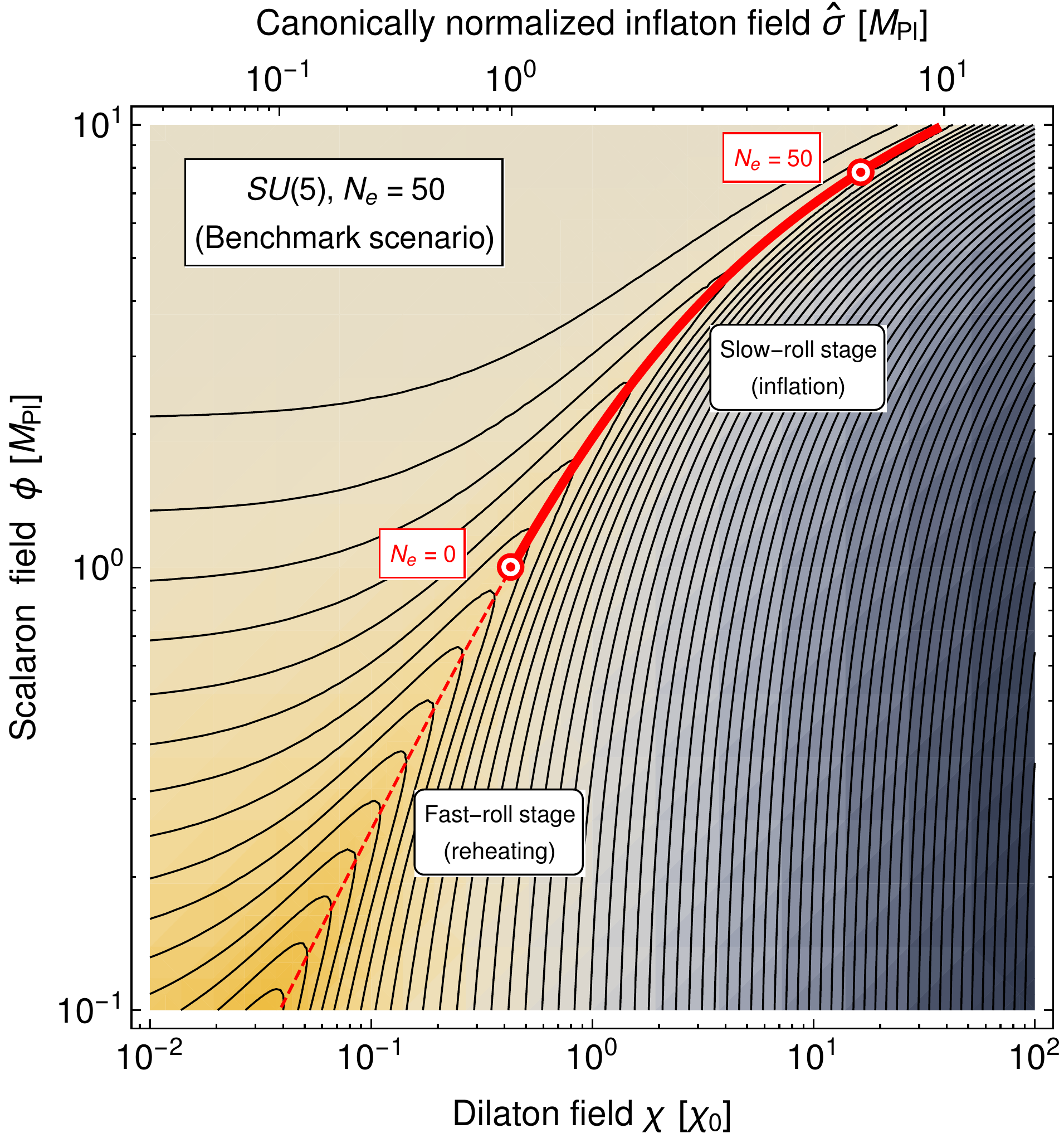}\hfill
\includegraphics[width=0.46\textwidth]{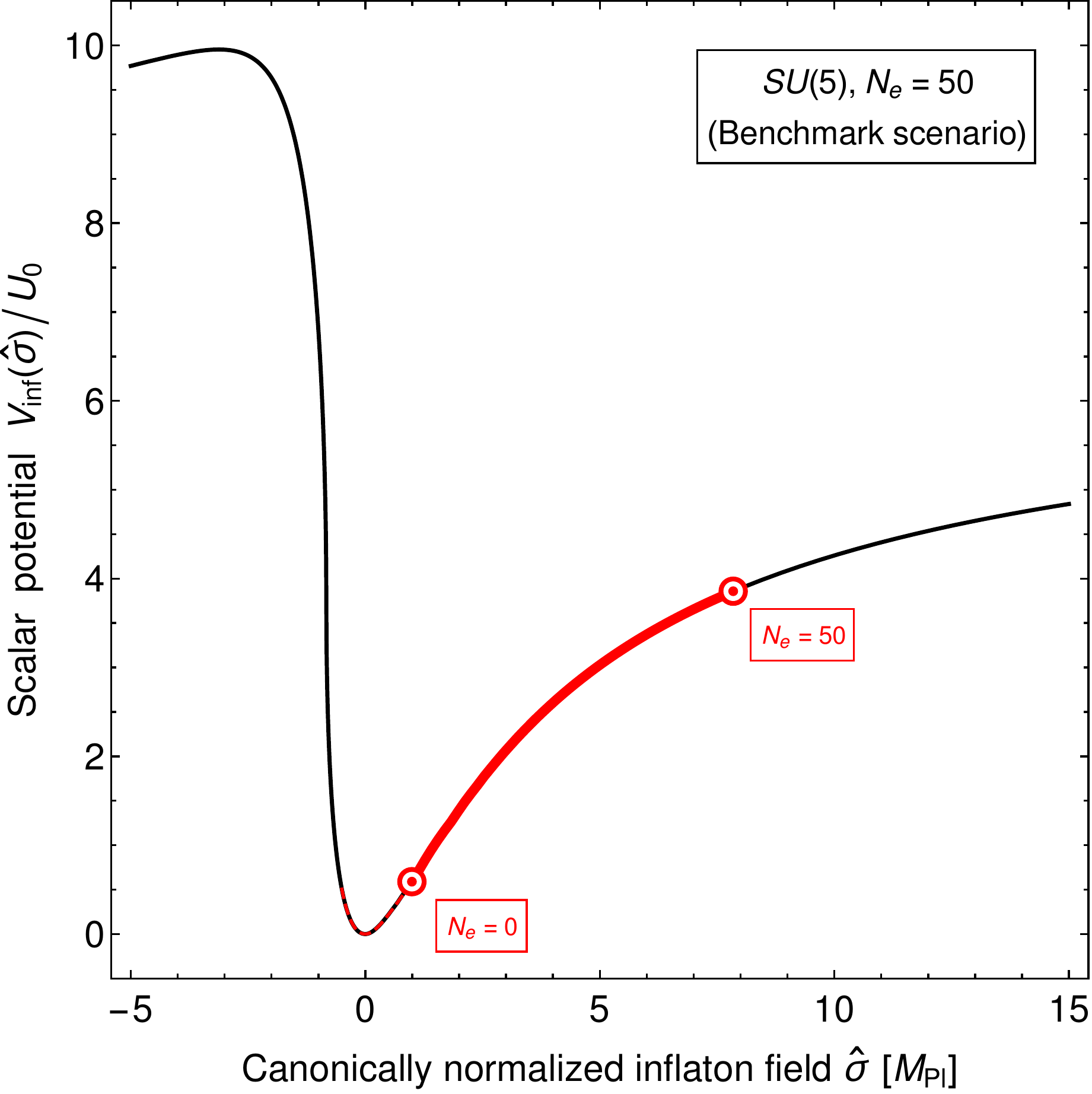}
\caption{Inflationary trajectory in field space \textbf{(left panel)}
and effective scalar potential during inflation \textbf{(right panel)}
for a particular benchmark point in parameter space: $N_c = 5$, $\lambda = 1$,
$\gamma = \beta^2 \simeq 3.7 \times 10^7$.
These parameter values are chosen such that they reproduce the correct values
for the amplitude and the spectral tilt of the scalar power spectrum.
The black contour lines in the left plot represent equipotential lines.
The color shading also indicates the height of the scalar potential.}

\label{fig:trajectory}
\end{center}
\end{figure}

Evaluating Eqs.~\eqref{eq:LEphichi} and \eqref{eq:V} along $\mathcal{C}$,
we obtain the following Lagrangian,
\begin{align}
\label{eq:Lsigma}
\left.\frac{\mathcal{L}_{\rm eff}^E}{\sqrt{- g}}\right|_{\mathcal{C}} =
- \frac{1}{2}\,M_{\rm Pl}^2\,R
+ \frac{1}{2}\,N_\sigma^2\left(\sigma\right)\,g^{\mu\nu}
\,\partial_\mu\sigma\,\partial_\nu\sigma
- V_{\rm inf}\left(\sigma\right) \,.
\end{align}
Note that the field $\sigma$ is not canonically normalized;
its kinetic term is multiplied by the field-dependent function $N_\sigma$,
\begin{align}
\label{eq:Nsigma}
N_\sigma\left(\sigma\right) = \frac{1}{\left(1+4\,A\left(\sigma\right)\right)B\left(\sigma\right)}
\left[\left(1+4\,A\left(\sigma\right)\right)B\left(\sigma\right)
+ \frac{3}{2}\,M_{\rm Pl}^2
\left(\left(1+4\,A\left(\sigma\right)\right)B'\left(\sigma\right)
+ 4\,A'\left(\sigma\right)B\left(\sigma\right)\right)^2\right]^{1/2} \,.
\end{align}
Meanwhile, the scalar potential for the noncanonically normalized inflaton field $\sigma$
obtains the following form,
\begin{align}
\label{eq:Vinf}
V_{\rm inf}\left(\sigma\right) = 
\frac{U\left(\sigma\right)}{\left(1 + 4\,A\left(\sigma\right)\right)
B^2\left(\sigma\right)} \,.
\end{align}
The canonically normalized inflaton field $\hat{\sigma}$ follows
from integrating the function $N_\sigma$ over an appropriate field range,
\begin{align}
\label{eq:sigmahat}
\hat{\sigma}\left(\sigma\right) = \int_0^\sigma d\sigma'\,N_\sigma\left(\sigma'\right) \,.
\end{align}
The relation between the three scalar fields $\chi$, $\phi$, and $\hat{\sigma}$
is illustrated in Fig.~\ref{fig:trajectory} for a particular
benchmark point in parameter space.
The left panel of this figure shows a contour plot of the scalar
potential $V$ in Eq.~\eqref{eq:V} as a function of $\chi$ and $\phi$,
together with the inflationary trajectory $\mathcal{C}$ in Eq.~\eqref{eq:contour},
while in the right panel of this figure, we plot the effective potential
during inflation in Eq.~\eqref{eq:Vinf}, $V_{\rm inf}$, as a function of $\hat{\sigma}$.


Let us now discuss the various conditions that need to be satisfied
so that the effective Lagrangian in Eq.~\eqref{eq:Lsigma} can be used
as a starting point for the description of inflation.
(i) A first obvious condition is that inflation can only occur along the contour
$\mathcal{C}$ if the extremum in the scalaron direction
corresponds to a local minimum.
In this case, inflation will proceed along a slightly sloped valley.
Otherwise, i.e., if the extremum corresponds to a local maximum,
the contour $\mathcal{C}$ will describe an unstable ridge in the potential landscape.
Therefore, a necessary condition for successful inflation is that, along the
trajectory $\mathcal{C}$, the scalaron direction is always
stabilized by a positive mass squared,
\begin{align}
\label{eq:mphi2}
m_\phi^2\left(\sigma\right) =
\left.\frac{\partial^2V}{\partial\phi^2}\right|_{\mathcal{C}}
> 0 \,.
\end{align}
Making use of the function $A$ introduced in Eq.~\eqref{eq:Adef},
the scalaron mass during inflation can be written as
\begin{align}
\label{eq:mphi2G}
m_\phi^2\left(\sigma\right) = \frac{M_{\rm Pl}^2}{12\left(1 + 4\,A\left(\sigma\right)
\right)G\left(\sigma\right)} \,.
\end{align}
Successful inflation requires a positive vacuum energy density, $V_{\rm inf} > 0$.
The expression for the scalar potential in Eq.~\eqref{eq:Vinf} thus implies
that $A$ must always be larger than $-1/4$.
For this reason, the condition of a positive scalaron mass
squared in Eq.~\eqref{eq:mphi2} translates into the statement 
that the function $G$ must always be positive during inflation,%
\footnote{$G>0$ implies in turn that $A$ cannot become negative.
As a matter of fact, $A$ thus never obtains values in the interval $-1/4 < A < 0$.}
\begin{align}
G\left(\sigma\right) = \bar{\gamma} - \frac{N_c\beta^2}{32\pi^2}
\,\ln\left(1 + \frac{\sigma}{\chi_0}\right)^2 > 0 \,.
\end{align}
We remark that the radiative corrections to the Lagrangian parameter $\gamma$
that are encoded in the function $G$ always come with a negative sign.
As a consequence, 
it is, in fact, impossible to satisfy the condition $G>0$ for all values
of the scalar field $\sigma$.
As soon as $\sigma$ exceeds a certain critical value $\sigma_{\rm crit}$,
the function $G$ turns negative,
\begin{align}
\sigma > \sigma_{\rm crit} = \left[\exp\left(\frac{16\pi^2\,\bar{\gamma}}
{N_c\beta^2}\right) - 1\right] \chi_0
\qquad\Rightarrow\qquad G\left(\sigma\right) < 0 \,.
\end{align}
However, for all practical purposes, it is sufficient if this critical field value
is large enough compared to the typical field values during inflation.
In addition, it may well be that the scalar potential receives
further gravitational corrections at field values far above the Planck
scale that regularize the singularity at $G = 0$.
For these reasons, we content ourselves with the requirement
that $\sigma_{\rm crit}$ should be at least as large as the Planck scale,
\begin{align}
\label{eq:sigmacrit}
\sigma_{\rm crit} = \left[\exp\left(\frac{16\pi^2\,\bar{\gamma}}
{N_c\beta^2}\right) - 1\right] \chi_0 > M_{\rm Pl} \,.
\end{align}
This condition ensures that the available field range in the $\sigma$
direction will always be large enough to support inflation.


(ii) Merely requiring $m_\phi^2$ to be positive is not enough to sufficiently
stabilize the scalaron at $\phi = \phi_\star$.
In fact, from the perspective of slow-roll inflation, we can only ignore
the motion in the scalaron direction away from the trajectory $\mathcal{C}$
if the scalaron mass $m_\phi$ is large compared to the
inflationary Hubble rate $H_{\rm inf}$ for all times during inflation,
\begin{align}
\label{eq:mphiH}
\frac{m_\phi^2\left(\sigma\right)}{H_{\rm inf}^2\left(\sigma\right)} =
\frac{1}{A\left(\sigma\right)} \gg 1 \,.
\end{align}
Here, we used the standard expression for $H_{\rm inf}$
in the slow-roll approximation,
\begin{align}
\label{eq:Hinf}
H_{\rm inf}\left(\sigma\right) =
\frac{V_{\rm inf}^{1/2}\left(\sigma\right)}{\sqrt{3}\,M_{\rm Pl}} =
\frac{U^{1/2}\left(\sigma\right)}{\sqrt{3}
\left(1 + 4\,A\left(\sigma\right)\right)^{1/2}B\left(\sigma\right)M_{\rm Pl}} \,.
\end{align}
In view of Eq.~\eqref{eq:mphiH}, we therefore conclude that the
function $A$ must always remain small during inflation, $A \ll 1$.
The inflationary Hubble rate $H_{\rm inf}$ in Eq.~\eqref{eq:Hinf} also determines the curvature scalar during slow-roll inflation, $\langle R\rangle \simeq 12\,H_{\rm inf}^2$.
One can show that, for typical field and parameter values during inflation, $\langle R\rangle$ is therefore always much smaller than the expectation value of the mean field in Sec.~\ref{sec:model}, $\langle R \rangle \ll \langle f \rangle$.
This justifies our assumption in deriving Eq.~\eqref{f0}.


Once the condition $A\ll 1$ becomes violated, we can no longer fully trust our ansatz
in Eqs.~\eqref{eq:phi0} and \eqref{eq:contour} and be sure that inflation
will indeed proceed along the contour $\mathcal{C}$.
In this case, we may alternatively look for local
minima in the dilaton direction $\chi$ and attempt to construct
an inflationary trajectory $\mathcal{C}'$ along these minima,
\begin{align}
\label{eq:Cprime}
\left.\frac{\partial\,V}{\partial\chi}\right|_{\left(\chi_\star\left(\phi\right),\phi\right)} = 0
\qquad\Rightarrow\qquad
\mathcal{C}'\left(\sigma\right) = 
\left\{\chi\left(\sigma\right),\phi\left(\sigma\right)\right\} = 
\left\{\chi_\star\left(\sigma\right),\sigma\right\} \,.
\end{align}
However, this approach is complicated by the fact that the form
of the scalar potential in Eq.~\eqref{eq:V} does not allow us to compute
the local minima in the dilaton direction analytically.
The best we can do in order to determine the function $\chi_\star$ is to adopt
an iterative procedure.
First, let us ignore for a moment the field dependence of all logarithmic
terms in Eq.~\eqref{eq:V}.
In this approximation, we are able to solve the extremal problem in Eq.~\eqref{eq:Cprime}
exactly.
Then, in the next step, we can use the result of the first step to find
a better approximation for the logarithmic terms, which provides us in turn with
a slightly improved solution for $\chi_\star$.
After $n$ iterations, this leads to the expression for $\chi_\star$,
\begin{align}
\label{eq:xstarn}
x_\star^{(n+1)}\left(\phi\right) = 1 + \frac{\chi_\star^{(n+1)}\left(\phi\right)}{\chi_0} = 
\left[1+2\,\ell\,L^{(n)}\left(\phi\right) 
- \frac{1-2\,L^{(n)}\left(\phi\right)}{1+2\,\ell\,L^{(n)}\left(\phi\right)}
\frac{16\,U_0}{M_{\rm Pl}^4}\,G^{(n)}\left(\phi\right)
\right]^{-1/2} e^{\Phi\left(\phi\right)/2} \,,
\end{align}
with the logarithmic terms after $n$ iterations, $L^{(n)}$ and $G^{(n)}$,
being defined as follows,
\begin{align}
L^{(n)}\left(\phi\right) = \ln \left(x_\star^{(n)}\left(\phi\right)\right)^2 \,, \quad 
G^{(n)}\left(\phi\right) = \bar{\gamma} - \frac{N_c\beta^2}{32\pi^2}\,
L^{(n)}\left(\phi\right)  \,. 
\end{align}
The initial $x_\star$ value is arbitrary and must be chosen
by hand.
For definiteness, we may, e.g., choose $x_\star^{(0)} = \sqrt{e}$ such that 
$L^{(0)} = 1$.
In an explicit numerical analysis, we are able to confirm that this
iterative procedure performs very well.
Over a broad range of parameter and field values, Eq.~\eqref{eq:xstarn}
manages to approximate the exact numerical result for $x_\star$ to high precision.
Moreover, we are able to show that, for $A \ll 1$, the contours $\mathcal{C}$
and $\mathcal{C}'$ always lie very close together in field space.
In this case, both approaches lead to the same inflationary dynamics
and hence to identical predictions for the primordial power spectra.
But on top of that, we are also able to confirm numerically
that, even for $A \gtrsim 1$, inflation along the contours $\mathcal{C}$
and $\mathcal{C}'$ leads to very similar predictions.
For this reason, we will from now on only focus on inflation
along the trajectory $\mathcal{C}$.
In the regions of parameter space where $A \gtrsim 1$, this is still
a sufficiently good approximation.
A dedicated numerical analysis of inflation along the contour
$\mathcal{C}'$ is left for future work. 


(iii) The fact that the dilaton-scalaron system can be effectively
described by a simple single-field model is also apparent from the
scalar mass spectrum.
After diagonalizing the dilaton-scalaron mass matrix and identifying the
two mass eigenstates, one typically finds that one mass
eigenvalue is extremely large.
It is therefore well justified to integrate out the heavy DOF
and describe inflation in terms of a single-field model.
To see this explicitly, one requires knowledge of the scalar mass
matrix along the contour $\mathcal{C}$.
The mass squared of the scalaron field is given in Eq.~\eqref{eq:mphi2G}.
The mass squared of the canonically normalized dilaton
field $\hat{\chi}$ as well as the dilaton-scalaron mass mixing
parameter can be calculated as follows,
\begin{align}
\label{eq:masses}
m_{\hat{\chi}}^2\left(\sigma\right) = \frac{1}{N_\chi^2\left(\sigma\right)}
\left.\frac{\partial^2V}{\partial\chi^2}\right|_{\mathcal{C}} \,, \quad
m_{\hat{\chi}\phi}^2\left(\sigma\right) = \frac{1}{N_\chi\left(\sigma\right)}
\left.\frac{\partial^2V}{\partial\chi\partial\phi}\right|_{\mathcal{C}} \,.
\end{align}
Here, we accounted for the noncanonical normalization of the dilaton field
in Eq.~\eqref{eq:LEphichi} by means of the factor $N_\chi$,
\begin{align}
d\hat{\chi} = N_\chi\left(\sigma\right) d\chi \,, \quad
N_\chi\left(\sigma\right) = e^{-\Phi\left(\phi_\star\left(\sigma\right)\right)/2} =
\frac{1}{\left[\left(1 + 4\,A\left(\sigma\right)\right)
B\left(\sigma\right)\right]^{1/2}} \,.
\end{align}
For simplicity, we shall neglect the field dependence of $N_\chi$ for the time being
and simply work with a symmetric mass matrix,
$m_{\phi\hat{\chi}}^2 = m_{\hat{\chi}\phi}^2$, in the following.
An explicit computation of the partial derivatives in Eq.~\eqref{eq:masses} results in
\begin{align}
\frac{m_{\hat{\chi}}^2\left(\sigma\right)}{N_\chi^2\left(\sigma\right)} & = 
U''\left(\sigma\right) + 2\left(
\frac{2\,B'\left(\sigma\right)G'\left(\sigma\right)}{B\left(\sigma\right)G\left(\sigma\right)}
- \frac{B''\left(\sigma\right)}{B\left(\sigma\right)}\right)U\left(\sigma\right)
+ 16\left(\frac{2\,G'^2\left(\sigma\right)}{B^2\left(\sigma\right) G\left(\sigma\right)}
- \frac{G''\left(\sigma\right)}{B^2\left(\sigma\right)}\right)
\frac{U^2\left(\sigma\right)}{M_{\rm Pl}^4}
+ \frac{B'^2\left(\sigma\right)}{8\,G\left(\sigma\right)}\,M_{\rm Pl}^4 \,,
\nonumber \\
\frac{m_{\hat{\chi}\phi}^2\left(\sigma\right)}{N_\chi^3\left(\sigma\right)} & =
- \frac{\sqrt{8}}{\sqrt{3}\,M_{\rm Pl}}
\left[U'\left(\sigma\right) + \left(\frac{G'\left(\sigma\right)}{G\left(\sigma\right)}
- \frac{B'\left(\sigma\right)}{B\left(\sigma\right)}\right) U\left(\sigma\right)
+ \frac{B\left(\sigma\right)B'\left(\sigma\right)}
{16\,G\left(\sigma\right)}\,M_{\rm Pl}^4\right] \,.
\end{align}
Together, the mass parameters $m_\phi^2$, $m_{\hat{\chi}}^2$, and $m_{\hat{\chi}\phi}^2$
can be used to determine the scalar mass eigenvalues $m_\pm^2$ during inflation.
Typically, we find one heavy mass eigenstate, $0 < H_{\rm inf}^2 \ll m_+^2$, as
well as one light tachyonic mass eigenstate, $-H_{\rm inf}^2 \ll m_-^2 < 0$.
The tachyonic mass eigenstate appears as a consequence of the negative curvature
of the scalar potential in the direction of the inflationary trajectory.
Similarly, the heavy mass eigenstate reflects the fact that the scalar potential
is steeply rising in the direction orthogonal to the contour $\mathcal{C}$.
In fact, an important condition for the validity of our analysis is that
the mass eigenvalue $m_+$ never exceeds the Planck scale,
\begin{align}
\label{eq:mplus}
m_+\left(\sigma\right) \lesssim M_{\rm Pl} \,.
\end{align}
Otherwise, we would no longer be able to trust our standard quantum field
theory analysis.
In our numerical study of slow-roll inflation (see Sec.~\ref{subsec:numerics}),
we explicitly compute $m_\pm$ along the inflationary trajectory at each point in parameter
space and check whether the condition in Eq.~\eqref{eq:mplus} is always fulfilled.
As it turns out, $m_+$ sometimes does become as large as $\mathcal{O}\left(M_{\rm Pl}\right)$,
but the bound in Eq.~\eqref{eq:mplus} is never severely violated.
We thus argue that the mass spectrum of our model is
under control and that we can trust our analysis in all
relevant regions of parameter space.


(iv) The fourth and last condition for successful inflation is concerned
with the perturbativity of our model.
Recall that our construction is based on a strongly coupled
gauge theory at high energies.
This implies the risk that our effective single-field model at low energies
may inherit some strong-coupling effects that would render a standard
analysis in terms of perturbation theory invalid.
Therefore, it is essential to study the self-interactions of the
canonically normalized inflaton field $\hat{\sigma}$ and examine whether
or not they remain in the perturbative regime.
The brute-force approach to this problem would be to explicitly compute the
scalar potential $V_{\rm inf}$ as a function of $\hat{\sigma}$ and to study its
properties in the most general terms.
However, to do so, one would have to invert the relation
between $\sigma$ and $\hat{\sigma}$ in Eq.~\eqref{eq:sigmahat}, which,
in practice, can only be done numerically.
For this reason, we will adopt a different approach in the following
that bypasses this technical complication.
The crucial point is that we are mostly interested in the properties of the
scalar potential in the vicinity of the origin in field space, where we
expect the inflaton self-interactions to be the strongest (see Fig.~\ref{fig:trajectory}).
For small field values, we are then able to expand
$V_{\rm inf}$ in powers of the inflaton field,
\begin{align}
V_{\rm inf}\left(\sigma\left(\hat{\sigma}\right)\right) =
\frac{1}{2}\,\hat{a}_2\,\hat{\sigma}^2
+ \frac{1}{3}\,\hat{a}_3\,\hat{\sigma}^3 + \frac{1}{4}\,\hat{a}_4\,\hat{\sigma}^4
+ \mathcal{O}\left(\hat{\sigma}^5\right) \,,
\end{align}
where we made use of the fact that the origin corresponds to a global Minkowski
vacuum, $V_{\rm inf}\left(0\right) = V_{\rm inf}'\left(0\right) = 0$, by construction.
Thanks to the chain rule, the coefficients in this Taylor expansion can be computed
solely in terms of the known expressions for $V_{\rm inf}$ and $N_\sigma$ as
functions of the noncanonically normalized inflaton field $\sigma$,
\begin{align}
\hat{a}_2 & = \left.\frac{V_{\rm inf}''\left(\sigma\right)}
{N_\sigma^2\left(\sigma\right)}\right|_{\sigma = 0} \,,
\\ \nonumber
\hat{a}_3 & = \frac{1}{2\,N_\sigma^3\left(\sigma\right)} \left[
V_{\rm inf}'''\left(\sigma\right) - 3\,\frac{N_\sigma'\left(\sigma\right)}
{N_\sigma\left(\sigma\right)}
\,V_{\rm inf}''\left(\sigma\right)\right]_{\sigma = 0} \,,
\\ \nonumber
\hat{a}_4 & = \frac{1}{6\,N_\sigma^4\left(\sigma\right)} \left[
V_{\rm inf}''''\left(\sigma\right)
- 6\,\frac{N_\sigma'\left(\sigma\right)}{N_\sigma\left(\sigma\right)}\,
V_{\rm inf}'''\left(\sigma\right)
+ 15\,\frac{N_\sigma'^2\left(\sigma\right)}{N_\sigma^2\left(\sigma\right)}\,
V_{\rm inf}''\left(\sigma\right)
- 4\,\frac{N_\sigma''\left(\sigma\right)}{N_\sigma\left(\sigma\right)}\,
V_{\rm inf}''\left(\sigma\right)\right]_{\sigma = 0} \,.
\end{align}
Based on these results, one can show that the self-interactions of the field
$\hat{\sigma}$ remain perturbative for all parameter values of interest.
On the analytical side, it is easy to see that the overall magnitude of the
coefficients is universally controlled by $U_0/M_{\rm Pl}^4$, i.e., the normalization
of the Jordan-frame scalar potential in units of the Planck mass,
\begin{align}
\frac{U_0}{M_{\rm Pl}^4} = \frac{N_c\,\lambda^2}{64\pi^2\,\beta^2} \,.
\end{align}
As long as this parameter is small, the coefficients $\hat{a}_{2,3,4}$
automatically obtain small values in Planck units.
With regard to the quartic self-coupling $\hat{a}_4$, this means
in particular that a sub-Planckian potential energy density in the Jordan frame,
$U_0 \ll M_{\rm Pl}^4$, readily results in
perturbatively small values, $\hat{a}_4 \ll 1$.
In our numerical slow-roll analysis, we are able to confirm these results
by explicitly evaluating the coefficients $\hat{a}_{2,3,4}$ at
each point in parameter space.


\subsection{Numerical analysis of the slow-roll dynamics}
\label{subsec:numerics}

\begin{figure}
\begin{center}
\includegraphics[width=0.475\textwidth]{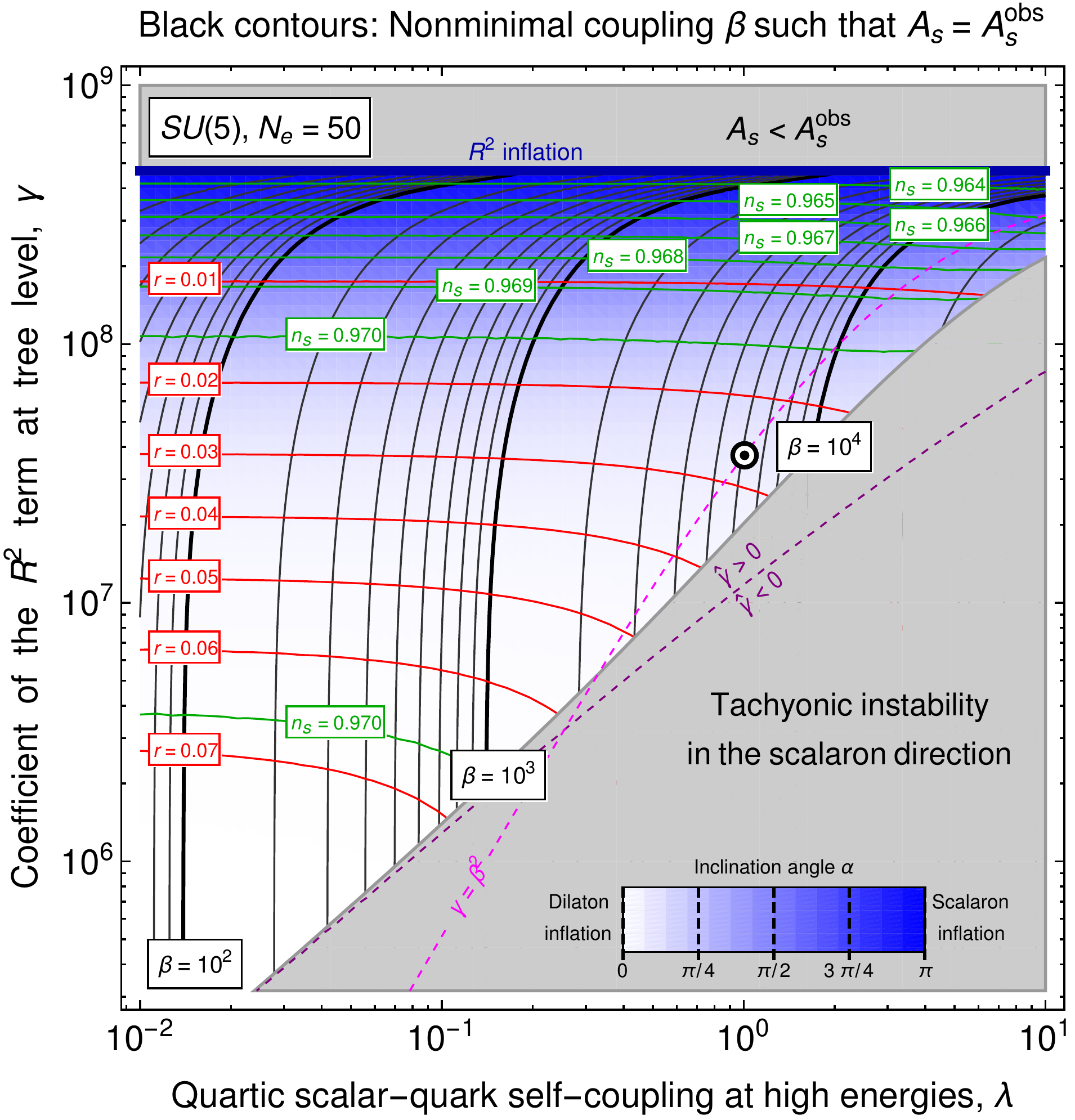}
\includegraphics[width=0.485\textwidth]{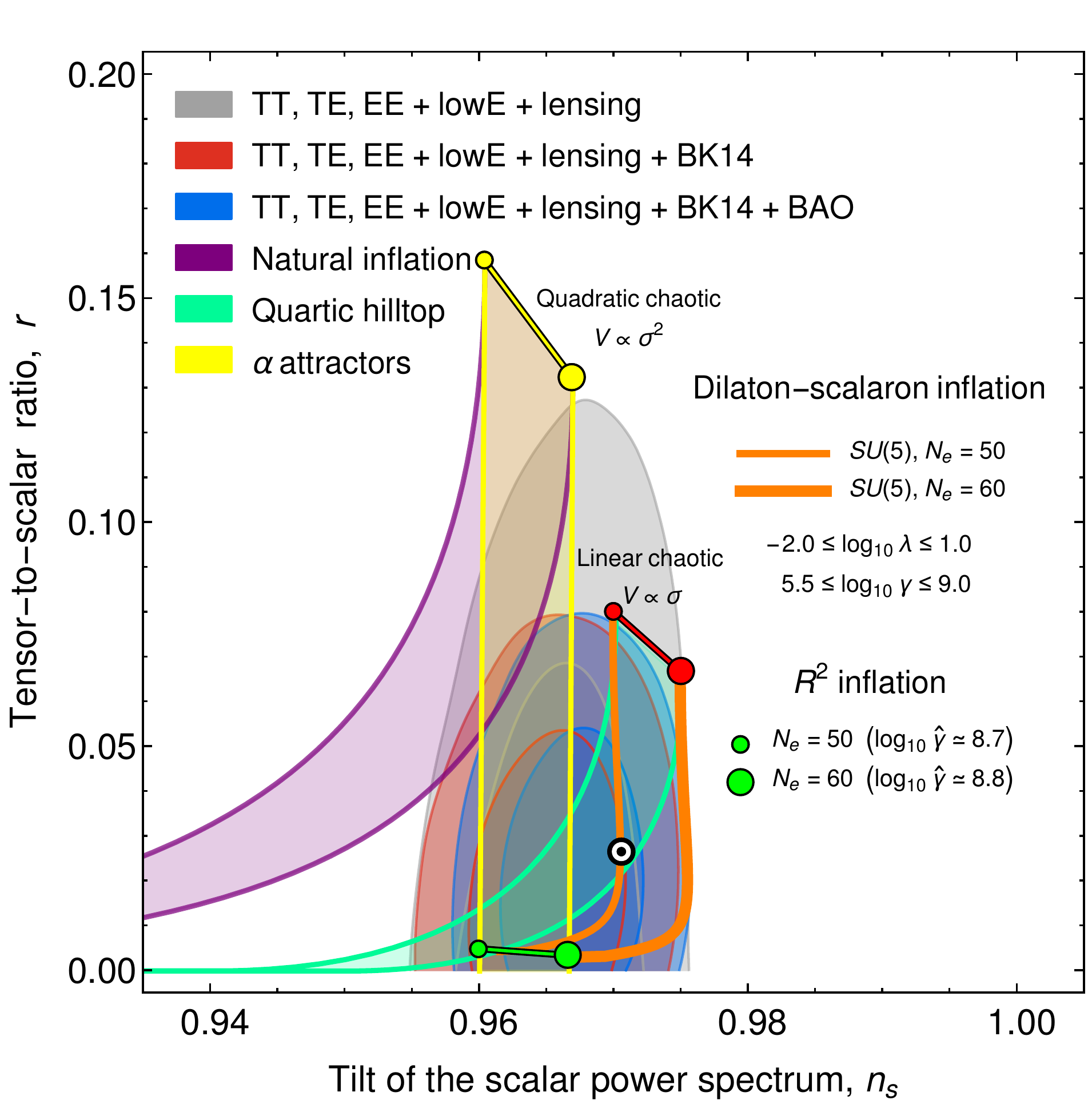}
\caption{Predictions for the CMB observables $n_s$ and $r$ in the parameter
space of our model \textbf{(left panel)} as well as in the $n_s$\,--\,$r$ plane
\textbf{(right panel)}.
The gray-shaded area in the lower part of the left plot is excluded because it leads
to a violation of the theoretical constraint in Eq.~\eqref{eq:sigmacrit}.
The inclination angle $\alpha$ is defined in Eq.~\eqref{eq:alpha}.
The PLANCK constraints in the right plot are taken from~\cite{Akrami:2018odb}.
For more details on the various reference models of inflation, see the
discussion and analysis in~\cite{Akrami:2018odb}.
The black and white circle in both plots corresponds to the
benchmark parameter point chosen in Fig.~\ref{fig:trajectory}.}
\label{fig:predictions}
\end{center}
\end{figure}

The effective single-field model specified in Eqs.~\eqref{eq:Lsigma}, \eqref{eq:Nsigma},
and \eqref{eq:Vinf} allows us to perform a standard single-field slow-roll analysis
of inflation.
The goal of this analysis is to compute the usual observables related to
the primordial power spectra encoded in the temperature fluctuations of
the CMB: the amplitude $A_s$ and the spectral tilt $n_s$
of the scalar power spectrum as well as the tensor-to-scalar ratio $r$.
These three observables are measured or constrained by the
latest data from the PLANCK satellite mission~\cite{Akrami:2018odb}.
At $68\,\%$\,C.\,L., the scalar amplitude $A_s$ is measured to be
\begin{align}
A_s = e^{3.044 \pm 0.014}  \times 10^{-10} \qquad
\textrm{(TT $+$ TE $+$ EE $+$ lowE $+$ lensing)}\,.
\end{align}
In the following, we will work with the best-fit value, $A_s^{\rm obs} =
e^{3.044} \times 10^{-10} \simeq 2.1 \times 10^{-9}$, for definiteness.
The combined constraints on $n_s$ and $r$ for various combinations
of datasets are shown in the right panel of Fig.~\ref{fig:predictions}.


Our predictions for $A_s$, $n_s$, and $r$ can be conveniently computed
in terms of the slow-roll parameters $\varepsilon$ and $\eta$,
\begin{align}
\label{eq:obs}
A_s = \frac{V_{\rm inf}}{24\pi^2\,\varepsilon\,M_{\rm Pl}^4} \,, \quad
n_s = 1 + 2\,\eta - 6\,\varepsilon \,, \quad
r = 16\,\varepsilon \,,
\end{align}
where all field-dependent quantities ($V_{\rm inf}$, $\varepsilon$, and $\eta$)
are supposed to be evaluated at the time of CMB horizon exit,
i.e., $N_e \simeq 50\cdots60$ $e$-folds before the end of inflation.
The parameters $\varepsilon$ and $\eta$ are defined in terms of partial
derivatives of the scalar potential w.r.t.\ the canonically normalized
inflaton field $\hat{\sigma}$.
However, thanks to the chain rule, these derivatives can again be readily
expressed as derivatives w.r.t.\ the noncanonically normalized
field $\sigma$,
\begin{align}
\label{eq:epseta}
\varepsilon\left(\sigma\right) & =
\frac{M_{\rm Pl}^2}{2}
\left(\frac{\partial_{\hat{\sigma}}
V_{\rm inf}\left(\sigma\left(\hat{\sigma}\right)\right)}
{V_{\rm inf}\left(\sigma\left(\hat{\sigma}\right)\right)}\right)^2 = 
\frac{M_{\rm Pl}^2}{2\,N_\sigma^2\left(\sigma\right)}
\left(\frac{\partial_\sigma V_{\rm inf}
\left(\sigma\right)}{V_{\rm inf}\left(\sigma\right)}\right)^2 \,,
\\ \nonumber
\eta\left(\sigma\right) & =
M_{\rm Pl}^2\,\frac{\partial_{\hat{\sigma}}\partial_{\hat{\sigma}}
V_{\rm inf}\left(\sigma\left(\hat{\sigma}\right)\right)}
{V_{\rm inf}\left(\sigma\left(\hat{\sigma}\right)\right)} =
\frac{M_{\rm Pl}^2}{N_\sigma^2\left(\sigma\right)} \left(
\frac{\partial_\sigma\partial_\sigma
V_{\rm inf}\left(\sigma\right)}{V_{\rm inf}\left(\sigma\right)} -
\frac{\partial_\sigma N_\sigma\left(\sigma\right)}{N_\sigma\left(\sigma\right)}
\frac{\partial_\sigma V_{\rm inf}\left(\sigma\right)}{V_{\rm inf}\left(\sigma\right)}\right) \,.
\end{align}
The inflaton field value at the time of CMB horizon exit
can be computed by solving the slow-roll equation of motion,
\begin{align}
\label{eq:srEOM}
\frac{d\,\sigma}{dN_e} =
\frac{M_{\rm Pl}^2}{N_\sigma^2\left(\sigma\right)}
\frac{\partial_\sigma V_{\rm inf}\left(\sigma\right)}{V_{\rm inf}\left(\sigma\right)} \,,
\end{align}
with the field value at the end of inflation, $\sigma_0 = \sigma\left(N_e = 0\right)$,
defined by the requirement that
$\max\left\{\varepsilon\left(\sigma_0\right),\left|\eta\left(\sigma_0\right)\right|\right\} = 1$.


Together, the relations in Eqs.~\eqref{eq:obs}, \eqref{eq:epseta}, and \eqref{eq:srEOM}
provide us with the necessary tools to determine our 
theoretical predictions for $A_s$, $n_s$, and $r$.
Because of the complicated form of the functions $N_\sigma$ and $V_{\rm inf}$
in Eqs.~\eqref{eq:Nsigma} and \eqref{eq:Vinf}, this can only be done numerically\,---\,at
least as long as one intends not to make any simplifying assumptions.
Consequently, we will content ourselves with a purely numerical analysis in this paper.
Any analytical investigation based on further assumptions is left for future work.
The parameter space of our inflationary model is spanned by three dimensionless
coupling constants:
(i) the quartic self-coupling $\lambda$ of the scalar quark field $S$,
(ii) the strength $\beta$ of the nonminimal coupling between
the field $S$ and the Ricci scalar $R$,
and (iii) the coefficient $\gamma$ of the bare $R^2$ term before accounting
for any radiative corrections.
We also recall that all three parameters are understood to correspond to those
couplings that appear in the effective action for the constituent scalar field $S$ after
integrating out the strong $SU\left(N_c\right)$ gauge dynamics.
In the following, one of these parameters can be eliminated by requiring
that the amplitude of the scalar power spectrum obtains the correct value,
$A_s = A_s^{\rm obs}$.
For definiteness, we will take this parameter to be the nonminimal coupling
constant $\beta$.
This leaves us with a two-dimensional parameter space spanned by $\lambda$ and $\gamma$,
in which we can compute our predictions for $n_s$ and $r$.
The outcome of this analysis is shown in Fig.~\ref{fig:predictions}.


A remarkable feature of our results is that they include and extend the well-known
results of $R^2$ inflation~\cite{Starobinsky:1980te}.
This connection can be illustrated in the following intuitive way:
First of all, let us consider the orientation of the inflationary trajectory
$\mathcal{C}$ in the dilaton-scalaron field space at the time of CMB horizon exit.
At each point in parameter space, we would like to know the following:
Does inflaton take place in the scalaron or dilaton direction?
In other words, in view of the scalar potential
in Eq.~\eqref{eq:V}, are the inflationary dynamics determined by
the function $e^\Phi$ or by the functions $B$, $G$, and $U$?
A convenient way to answer these questions is to introduce the inclination
angle
\begin{align}
\label{eq:alpha}
\alpha = \arctan \left(\frac{\chi_0}{M_{\rm Pl}}
\frac{d\,\phi_\star}{d\chi}\right) \,,
\end{align}
which measures the angle between the dilaton axis in field space and 
the inflationary trajectory.%
\footnote{Here, we consider the
dimensionless quantities $\chi/\chi_0$ and $\phi/M_{\rm Pl}$ as the relevant
field variables in the scalar potential.
Only after this rescaling of the fields $\chi$ and $\phi$ are we able to compare 
the importance of the functions $e^\Phi$, $B$, $G$, and $U$ in a meaningful way.}
An inclination angle close to $0$ indicates that the dynamics of inflation
are mostly governed by the dilaton-dependent functions in the scalar potential.
Conversely, an inclination angle close to $\pi$ points to inflation being dominated
by the scalaron-dependent part of the scalar potential.
In Fig.~\ref{fig:predictions}, we refer to these two scenarios as dilaton
and scalaron inflation, respectively.
In particular, we indicate the dependence of $\alpha$ as a function of $\lambda$
and $\gamma$ by the color shading in the left panel of Fig.~\ref{fig:predictions}.


Interestingly enough, $\alpha$ converges towards $\alpha \simeq \pi$ as soon as
$\gamma$ approaches a characteristic value of $\mathcal{O}\left(10^9\right)$.
This is nothing but the $R^2$-inflation limit of our model.
In this limit, the dynamics of the dilaton field become increasingly
irrelevant from the perspective of inflation.
In fact, for large values of $\gamma$, the value of $\beta$ that is required
to obtain the correct scalar amplitude decreases.
According to Eq.~\eqref{eq:B}, this results in a large value of $\chi_0$,
such that
\begin{align}
\chi \ll \chi_0 \qquad\Rightarrow\qquad
B\left(\chi\right) \approx 1 \,, \quad
G\left(\chi\right) \approx \bar{\gamma} \,, \quad
U\left(\chi\right) \approx 0 \,.
\end{align}
As a consequence, the scalar potential in Eq.~\eqref{eq:V} approaches
the well-known potential of $R^2$ inflation~\cite{Starobinsky:1980te},
\begin{align}
\chi \ll \chi_0 \qquad\Rightarrow\qquad
V\left(\chi,\phi\right) \approx \frac{M_{\rm Pl}^4}{16\,\bar{\gamma}}\left(1
- e^{-\Phi\left(\phi\right)}\right)^2 \,.
\end{align}
In this potential, we recognize the radiatively corrected parameter $\bar{\gamma}$
as the coefficient of the $R^2$ term in $R^2$ inflation.
The requirement of a correctly normalized scalar power spectrum then uniquely
fixes $\bar{\gamma}$ at a value of $\mathcal{O}\left(10^9\right)$,
\begin{align}
A_s = A_s^{\rm obs} \,,\quad 50 \leq N_e \leq 60 \qquad\Rightarrow\qquad
0.5 \times 10^9 \lesssim \bar{\gamma} \lesssim 0.7 \times 10^9 \,.
\end{align}
Up to terms suppressed by higher powers of $N_e^{-1}$, $R^2$ inflation
results in the following predictions for $n_s$ and $r$,
\begin{align}
n_s \simeq 1 - \frac{2}{N_e} \,, \quad r \simeq \frac{12}{N_e^2} \,.
\end{align}
For $N_e \simeq 50\cdots60$, this means that $r$
is expected to remain rather small, $3 \lesssim r/10^{-3}\lesssim 5$,
in the $R^2$-inflation limit.


However, this situation changes as soon as we make use of the parametric
freedom of our model.
For $\gamma$ values smaller than those required by $R^2$ inflation, the dilaton
dynamics begin to become important.
This helps us increase the tensor-to-scalar
ratio (see Fig.~\ref{fig:predictions}).
In fact, for small values of $\gamma$ and $\lambda$, the predictions
of our model approach those of linear chaotic inflation~\cite{Linde:1983gd}, which
is described by a scalar potential that is linear in the inflaton
field, $V_{\rm inf} \propto \sigma$.
This can be understood as follows:
For small values of $\gamma$ and $\lambda$, we may neglect
the function $A$ in Eqs.~\eqref{eq:Nsigma} and \eqref{eq:Vinf}.
In this approximation, it is possible to simplify
the scalar potential $V_{\rm inf}$
and the normalization factor $N_\sigma$ as follows,
\begin{align}
V_{\rm inf}\left(\sigma\right) \approx
\frac{U\left(\sigma\right)}{B^2\left(\sigma\right)} \,, \quad 
N_\sigma\left(\sigma\right) \approx \frac{1}{B\left(\sigma\right)}
\left[B\left(\sigma\right)
+ \frac{3}{2}\,M_{\rm Pl}^2\,B^{\prime\,2}\left(\sigma\right)\right]^{1/2} \,.
\end{align}
We are also able to neglect the logarithmic term
in the function $B$ in Eq.~\eqref{eq:B}, such that
$B\left(\sigma\right) \approx x^2\left(\sigma\right)$ and
\begin{align}
\label{eq:Vinfapprox}
V_{\rm inf}\left(\sigma\right) \approx
U_0 \left[\frac{1}{x^4\left(\sigma\right)} - 1 +
2\,\ln\left(x^2\left(\sigma\right)\right)\right] \,, \quad
N_\sigma\left(\sigma\right) \approx \left(1 + \frac{6\,M_{\rm Pl}^2}{\chi_0^2}\right)^{1/2}
\frac{1}{x\left(\sigma\right)} \,,
\end{align}
where $x$ is defined similarly as in Eq.~\eqref{eq:xstarn},
$x\left(\sigma\right) = 1 + \sigma/\chi_0$.
Given this form of the normalization factor $N_\sigma$ and making use of the
relation in Eq.~\eqref{eq:sigmahat}, we are able to determine
the canonically normalized inflaton field $\hat{\sigma}$,
\begin{align}
\hat{\sigma}\left(\sigma\right) \approx 2\,\sigma_0
\ln\left(x^2\left(\sigma\right)\right) \,, \quad
\sigma_0 = \frac{1}{4}\left(6\,M_{\rm Pl}^2 + \chi_0^2\right)^{1/2} \,.
\end{align}
Solving this relation for $\sigma$ and inserting the result
into $V_{\rm inf}$ in Eq.~\eqref{eq:Vinfapprox}
provides us with the scalar potential for $\hat{\sigma}$,
\begin{align}
\label{eq:Vinflin}
V_{\rm inf}\left(\sigma\left(\hat{\sigma}\right)\right) \approx 
U_0 \left[e^{-\hat{\sigma}/\sigma_0} - 1 + \frac{\hat{\sigma}}{\sigma_0}\right] \,.
\end{align}
At large field values, $\hat{\sigma} \gg \sigma_0$, this is a linear potential
just like in the case of linear chaotic inflation!
At the same time, the exponential correction regularizes the derivative
of the scalar potential around the origin in field space.
In fact, thanks to this exponential term, the scalar potential exhibits
a stable global minimum at $\hat{\sigma} = 0$.
Interestingly enough, a scalar potential similar to the one in
Eq.~\eqref{eq:Vinflin} can also be derived in the context of 
Coleman-Weinberg inflation supplemented by a nonminimal inflaton
coupling to gravity~\cite{Kannike:2015kda}
(see also~\cite{Barrie:2016rnv,Racioppi:2017spw,Racioppi:2018zoy}).
The scenario in~\cite{Kannike:2015kda} is, however,
distinct from ours in the sense that it employs weakly coupled
interactions to generate the Planck scale, whereas our model is
based on dimensional transmutation in a strongly coupled sector.
In addition, we find that our predictions for the inflationary CMB observables
approach those of linear chaotic inflation from below (see Fig.~\ref{fig:predictions}),
whereas in~\cite{Kannike:2015kda}, the predictions of linear chaotic inflation result
in a lower bound on the tensor-to-scalar ratio.


We now understand why small values of $\gamma$ and $\lambda$
lead to the same predictions as linear chaotic inflation,
\begin{align}
n_s  = 1 - \frac{3}{2\,N_e} \,, \quad r = \frac{4}{N_e} \,.
\end{align}
For $N_e = 50$, the tensor-to-scalar ratio can hence become as large as $r \simeq 0.08$,
while for $N_e = 60$, one can achieve values as large as $r \simeq 0.07$.
These are remarkable results, as they promise that our model can be probed
in future CMB experiments that are sensitive to $r$ values of the order
of $r \simeq \textrm{few} \times 10^{-2}$.
Moreover, it is important to note that our model populates
regions in the $n_s$\,--\,$r$ plane that are inaccessible in
other popular inflation models (see Fig.~\ref{fig:predictions}).


\section{Conclusions}


In this paper, we studied the dynamical generation of the Planck mass in
consequence of spontaneous conformal-symmetry breaking.
To this end, we constructed an explicit field-theoretical model based on the
idea that classical scale invariance becomes spontaneously broken
because of the strongly coupled gauge dynamics in a hidden sector.
Our construction led us to a remarkable realization:
\textit{The stage of cosmic inflation in the early Universe may be a consequence of the
microscopic physics that is responsible for the dynamical generation of the Planck mass!}
If correct, this observation would point to a
deep connection between particle physics and cosmology.
We emphasize that the pivotal element of our construction is the concept of classical
scale invariance\,---\,a new approximate global symmetry that forbids
any dimensionful coupling constant in the tree-level Lagrangian.
At the quantum level, scale invariance is explicitly broken
by the conformal anomaly, which reflects the logarithmic running of the
renormalized coupling constants.
However, as a working assumption, one may speculate that
this explicit breaking is eventually going to vanish at very high energies because
of the theory's renormalization group flow reaching an ultraviolet fixed point.%
\footnote{It has been suggested in~\cite{Wetterich:2014gaa}
that the infinite past (future) of the Universe may
correspond to an ultraviolet (infrared) fixed point.}


Within the paradigm of classical scale invariance, explicit mass
scales such as the Planck mass $M_{\rm Pl}$ can only be generated
radiatively by means of spontaneous symmetry breaking.
In our model, we follow this philosophy and show how classical scale invariance
may be spontaneously broken by means of dimensional transmutation in a hidden
strongly coupled $SU\left(N_c\right)$ gauge sector.
In particular, we consider a scalar bilinear operator $S^\dagger S$ that
is nonminimally coupled to the Ricci scalar and that condenses in
the confining phase of the $SU\left(N_c\right)$ gauge theory at low energies.
Of course, a similar construction making use of a
chiral condensate is also conceivable.
In this work, we focused on a scalar condensate for simplicity.
A study of the fermionic case is left for future work.
Up to a dimensionless coupling constant $\beta$, the vacuum expectation value
of the scalar condensate then directly determines the Planck mass,
\begin{align}
\label{eq:MPlSS}
\frac{M_{\rm Pl}^2}{2} = \beta \left<S^\dagger S\right> \,.
\end{align}
The vacuum expectation value of the scalar condensate
can be computed according to the following recipe:
\begin{enumerate}
\item Write down the most general action compatible with all global and local
symmetries of our model.
\item Construct an effective model for the constituent scalar field $S$
along the lines of the NJL model in QCD.
\item Apply the mean-field approximation to describe the dynamics
of the order parameter $f = S^\dagger S$.
\item Compute the quantum effective action for $f$ by integrating out
the heavy constituent scalar $S$.
\item Determine the symmetry-breaking minimum of the effective scalar potential
for the mean field $f$.
\end{enumerate}
This procedure results in a Coleman-Weinberg-like correction to the Lagrangian
of the gravitational sector,
\begin{align}
\label{eq:LeffCW}
\frac{\mathcal{L}_{\rm eff}}{\sqrt{-g}} \supset \gamma\,R^2 +
\frac{N_c}{32\pi^2}\,M_R^4\,\ln\frac{M_R^2}{\Lambda^2} \,, \quad
M_R^2 = 2\,\lambda\left<S^\dagger S\right> + \beta\,R \,.
\end{align}
For small values of $R$, the logarithmic term can be expanded in powers of $R$, which
generates among other terms the Einstein-Hilbert term in the gravitational action.
The coefficient of this term is then identified as the Planck mass.
An important aspect of this construction is that the gravitational scale
$M_{\rm Pl}$ originates from an ordinary energy scale in quantum field
theory, i.e., the vacuum expectation value of the scalar condensate
[see Eq.~\eqref{eq:MPlSS}].
Likewise, the absolute energy scale of the scalar potential after spontaneous
conformal-symmetry breaking, $U_0^{1/4}$,
is initially computed in field theory [see Eq.~\eqref{CC}].
Therefore, before going the next step and making contact with gravity, the potential
energy scale $U_0^{1/4}$ is expected to be subjected to appropriate
renormalization conditions on the field-theory side.
This observation may be useful in more realistic attempts at solving
the problem of the cosmological constant.


A remarkable consequence of our mechanism to generate the Planck mass
is that it automatically results in the presence of an additional scalar
DOF: the dilaton field $\chi$, which parametrizes the fluctuations
around the symmetry-breaking vacuum and plays the role of the
pseudo-NG boson of spontaneously broken scale invariance.
As we were able to show in the second part of this paper, this dilaton field
can be used to construct a successful model of inflation.
In fact, the dilaton couples to a second scalar DOF, the scalaron
field $\phi$, which originates from the $R^2$ term in the effective action.
A main result of our analysis is that the coupled dilaton-scalaron system
gives rise to an effective single-field model of inflation in the Einstein frame.
The predictions of this model for the inflationary CMB observables interpolate
between those of standard $R^2$ inflation and those of linear chaotic
inflation.
This important result represents a characteristic
phenomenological feature of our model.
As a consequence, we typically find values of the scalar spectral index
in the range $0.970 \lesssim n_s \lesssim 0.975$ and values for the
tensor-to-scalar ratio as large as $r \simeq 0.08$ (see Fig.~\ref{fig:predictions}).
These are exciting predictions that will be tested in future CMB experiments.


Our study presented in this paper should be understood as a first step towards
a better understanding of inflation in classically scale-invariant models.
There are certainly several directions into which our discussion may be extended:
(i) It would be worthwhile to study the full two-field dynamics 
of our model (see, e.g., \cite{Wands:2007bd,Buchmuller:2014epa})
and to investigate which other inflationary solutions one might be able to identify.
In this context, one should also compute the degree of non-Gaussianities generated
by our model and compare them to the current observational bounds.
(ii) One may refrain from expanding the effective scalar potential in powers of $R$
and keep working with the full expression in Eq.~\eqref{eq:LeffCW}.
This would amount to the study of a typical $f\left(R\right)$ model,
where $f\left(R\right)$ would schematically 
be given by~\cite{Rinaldi:2014gha,Liu:2018hno}
\begin{align}
f\left(R\right) = \gamma\,R^2 + V_{\rm CW}\left(R\right) \,.
\end{align}
(iii) Reheating after inflation deserves a dedicated analysis, in particular,
as this may allow to constrain the number of $e$-folds $N_e$
during inflation more precisely (see, e.g.,~\cite{Martin:2013tda}).
However, all of these questions are beyond the scope of this paper.
We leave them for future work, hoping that our analysis
will only be the first in a series of papers that address the possible connection
between the Planck mass, inflation, and the principle of classical scale invariance.


\bigskip\noindent\textit{Acknowledgments:}
The authors would like to thank Marco Drewes, Jean-Marc G\'erard, Kohei Kamada,
and Christophe Ringeval for useful comments and discussions.
K.\,S.\ is grateful to the Centre for Cosmology, Particle Physics, and Phenomenology
at the Universit\'e Catholique de Louvain for its hospitality during the
final stages of this project.
The work of M.\,Y.\ is supported by the DFG Collaborative Research
Centre SFB 1225 (ISOQUANT) and the Alexander von Humboldt foundation.
The work of J.\,K.\ is partially supported by the
Grant-in-Aid for Scientific Research (C) from the
Japan Society for the Promotion of Science (Grant No.\ 16K05315).


\bibliographystyle{JHEP}
\bibliography{arxiv_2}


\end{document}